\documentstyle[prd,preprint,tighten,aps,eqsecnum,%
amssymb,newlfont,epsfig]{revtex}
\begin{document}
\draft
\preprint{
\begin{tabular}{r}
UWThPh-1997-51\\
DFTT 74/97\\
hep-ph/9712537
\end{tabular}
}
\title{Long-baseline neutrino oscillation
experiments and CP violation in the lepton sector}
\author{S.M. Bilenky}
\address{Joint Institute for Nuclear Research, Dubna, Russia, and\\
INFN, Sezione di Torino, and Dipartimento di Fisica Teorica,
Universit\`a di Torino,\\
Via P. Giuria 1, I--10125 Torino, Italy}
\author{C. Giunti}
\address{INFN, Sezione di Torino, and Dipartimento di Fisica Teorica,
Universit\`a di Torino,\\
Via P. Giuria 1, I--10125 Torino, Italy}
\author{W. Grimus}
\address{Institute for Theoretical Physics, University of Vienna,\\
Boltzmanngasse 5, A--1090 Vienna, Austria}
\date{December 29, 1997}
\maketitle
\begin{abstract}
We discuss possibilities to investigate the effects of CP (and T)
violation in the lepton sector
in neutrino oscillation experiments.
We consider the effects of CP violation
in the framework of two schemes of mixing of four 
massive neutrinos that can accommodate the results of all
neutrino oscillation experiments.
Using the constraints on the mixing parameters
that follow from the results of
short-baseline neutrino oscillation experiments,
we derive rather strong upper bounds on the effects of CP violation in
$
\stackrel{\scriptscriptstyle(-)}{\nu}_{{\hskip-4pt}\mu} 
\leftrightarrows
\stackrel{\scriptscriptstyle(-)}{\nu}_{{\hskip-4pt}e}
$
transitions
in long-baseline
neutrino oscillation experiments.
We show that the effects of CP violation in
$
\stackrel{\scriptscriptstyle(-)}{\nu}_{{\hskip-4pt}\mu} 
\leftrightarrows
\stackrel{\scriptscriptstyle(-)}{\nu}_{{\hskip-4pt}\tau}
$
transitions in long-baseline oscillation experiments
can be as large as is allowed
by the unitarity of the mixing matrix.
The matter effects,
which complicate the problem of searching for
CP violation
in long-baseline experiments,
are discussed in detail.
We consider the T-odd asymmetries
whose measurement could allow
to reveal T and CP violation
in the lepton sector
independently from matter effects.
\end{abstract}


\narrowtext

\section{Introduction}
\label{Introduction}

The violation of CP invariance is one of
the most important problems in particle physics. So far, CP violation
has only been observed in the $K^0 \bar K^0$ system \cite{CPviolation}.
Many future experiments will investigate
the effects of CP violation in decays of B mesons and
are aimed
to reveal the origin of CP violation in the quark
sector (for recent reviews see, e.g., Ref. \cite{CPSM}). 
In the Standard Model of electroweak interactions CP violation resides
in a phase in the Cabibbo--Kobayashi--Maskawa mixing matrix \cite{KM}.
In the lepton sector
an analogous mixing matrix is expected to exist
if neutrinos are massive
and, consequently, it is plausible to assume the presence
of CP-violating phases in the neutrino mixing matrix as well.

At present
there are some indications
that neutrinos are massive and mixed particles
coming from the results of
solar neutrino experiments
(Homestake \cite{Homestake},
Kamiokande \cite{Kam-sol},
GALLEX \cite{GALLEX},
SAGE \cite{SAGE}
and
Super-Kamiokande \cite{SK-sol}),
of atmospheric neutrino experiments 
(Kamiokande \cite{Kam-atm},
IMB \cite{IMB},
Soudan \cite{Soudan}
and
Super-Kamiokande \cite{SK-atm})
and of the accelerator LSND experiment \cite{LSND}.
The analysis of the data of these experiments
in terms of neutrino oscillations
indicate the existence of three different
scales of neutrino mass-squared differences:
\begin{eqnarray}
&
\Delta{m}^2_{\mathrm{sun}}
\sim
10^{-5} \, \mathrm{eV}^2
\mbox{\cite{sol-msw}}
\quad \mbox{or} \quad
\Delta{m}^2_{\mathrm{sun}}
\sim
10^{-10} \, \mathrm{eV}^2
\mbox{\cite{sol-vac}}
\,,
&
\label{SUNrange}
\\
&
\Delta{m}^2_{\mathrm{atm}}
\simeq
5 \times 10^{-3} \, \mathrm{eV}^2
\mbox{\cite{SK-atm}}
\,,
&
\label{ATMrange}
\\
&
0.3 \lesssim \Delta{m}^2_{\mathrm{LSND}} \lesssim 2.2 \, \mathrm{eV}^2
\mbox{\cite{LSND}}
\,.
&
\label{LSNDrange}
\end{eqnarray}
The two possibilities for
$\Delta{m}^2_{\mathrm{sun}}$
correspond,
respectively,
to the
MSW \cite{MSW}
and
to the
vacuum oscillation
(see Refs. \cite{BP78,BP87,CWKim})
solutions of the solar neutrino problem.
At present
there is no information on CP violation
in the lepton sector.

Here we consider possibilities to reveal effects
of CP violation in neutrino oscillations
in schemes of neutrino mixing
that
can provide three independent
neutrino mass-squared differences.
These schemes
are based on the assumption that
the flavour neutrino fields are superpositions of four
massive neutrino fields.
This means that the neutrino mixing matrix
can contain CP-violating phases
and effects of CP violation in the lepton sector
could be observed in neutrino oscillation experiments.

In principle,
short-baseline (SBL) accelerator neutrino oscillation 
experiments could be 
important sources of information on CP-violation in 
the lepton sector,
but
in the case of four massive neutrinos
only the largest mass-squared difference
$\Delta{m}^2\equiv\Delta{m}^2_{\mathrm{LSND}}$
is relevant for SBL oscillations
and
the effects of CP violation cannot be revealed in SBL 
experiments
\cite{CPrelations}.
This fact is discussed
in Section \ref{Neutrino oscillations and CP violation},
where we review some general aspects of
CP violation in neutrino oscillations.
In the rest of the paper
we discuss the effects of CP violation
that can be expected
in future accelerator long-baseline (LBL) neutrino
oscillation experiments
(K2K \cite{K2K}, MINOS \cite{MINOS}, ICARUS \cite{ICARUS}
and others \cite{otherLBL}).

In Refs. \cite{BGG96A,BGG96B}
we have shown that
among all the possible schemes with four massive neutrinos
only two can accommodate the results of all
neutrino oscillation experiments (see also Ref.\cite{OY96}).
These two schemes are presented in
Section \ref{Four massive neutrinos}.
In
Section \ref{CP violation in the schemes with four neutrinos}
we apply to these schemes the general methods presented in the
Appendices \ref{apa}--\ref{apc}
that allow to obtain limits on 
the parameters that characterize
the CP-odd asymmetries in different LBL channels
from the exclusion plots obtained in SBL experiments.
We show that in the schemes under consideration
there are
rather severe constraints on the parameter
$I_{e\mu}$
that characterizes
CP violations in the
$
\stackrel{\scriptscriptstyle(-)}{\nu}_{{\hskip-4pt}\mu} 
\leftrightarrows
\stackrel{\scriptscriptstyle(-)}{\nu}_{{\hskip-4pt}e}
$
channels in vacuum,
whereas
the parameter
$I_{\mu\tau}$
that characterizes
CP violations in
$
\stackrel{\scriptscriptstyle(-)}{\nu}_{{\hskip-4pt}\mu} 
\leftrightarrows
\stackrel{\scriptscriptstyle(-)}{\nu}_{{\hskip-4pt}\tau}
$
LBL transitions in vacuum can reach its
maximum value determined by the unitarity of the mixing matrix.

The possible effects of CP violation
in LBL neutrino oscillation experiments
have been discussed recently in the literature
\cite{TA96,AR96,MN97}
in the framework of three-neutrino schemes
that could accommodate the results of some,
but not all,
neutrino oscillation experiments
(in particular, in these schemes the solar neutrino problem cannot
be explained with neutrino oscillations).
In Section \ref{Three massive neutrinos}
we apply the methods presented in this paper
in order to obtain the limits
on the parameters that characterize
the CP-odd asymmetries in different LBL channels
in the framework of these schemes of mixing of three massive neutrinos.

In Section \ref{Matter effects}
we discuss the implications of matter effects
for the possibility to observe CP violation in LBL experiments.
Matter effects can be large
and they represent a serious problem for the investigation
of CP violation in LBL experiments
because the interaction of neutrinos and antineutrinos 
with matter is not CP-symmetric.
In order to extract
the CP-violating phases of the mixing matrix
from the measured 
asymmetries it is necessary to have detailed information on 
the absolute values of the elements of the mixing matrix
and on the neutrino mass-squared differences.
To avoid this problem,
we consider
the T-odd asymmetries
whose measurement could reveal
CP violation
in the lepton sector independently from the presence of matter 
(the matter contribution to the effective Hamiltonian is T-symmetric)
\cite{KP87}.
Measurements of such asymmetries may be
possible in the future,
for example, with neutrino beams
from muon colliders
\cite{Geer,Mohapatra97}.

\section{Neutrino oscillations and CP violation}
\label{Neutrino oscillations and CP violation}

In accordance with the neutrino mixing hypothesis
(see, for example, Refs. \cite{BP78,BP87,CWKim}),
a left-handed neutrino field
$\nu_{{\alpha}L}$
is a mixture of the left-handed
components
$\nu_{kL}$
of the (Dirac or Majorana) fields of neutrinos
with definite masses
$m_k$:
\begin{equation}
\nu_{{\alpha}L}
=
\sum_{k}
U_{{\alpha}k}
\,
\nu_{kL}
\quad \mbox{with} \quad
\alpha=e,\mu,\tau,s,\ldots
\,,
\label{05}
\end{equation}
where $U$ is the unitary mixing matrix.
Here
$ k \geq 3 $
and
$ \nu_{sL} , \ldots $
are possible sterile neutrino fields.
The mixing in Eq.(\ref{05})
implies that
the transition probabilities in vacuum
of neutrinos and antineutrinos with momentum $p$ at a distance $L$ of
the neutrino detector from the neutrino source
are given by
\begin{eqnarray}
&&
P_{\nu_\alpha\to\nu_\beta}
=
\left|
\sum_{k}
U_{{\beta}k}
\,
U_{{\alpha}k}^{*}
\,
\exp\!\left(
- i \, \frac{ \Delta{m}^{2}_{k1} \, L }{ 2 \, p }
\right)
\right|^2
\,,
\label{052}
\\
&&
P_{\bar\nu_\alpha\to\bar\nu_\beta}
=
\left|
\sum_{k}
U_{{\beta}k}^{*}
\,
U_{{\alpha}k}
\,
\exp\!\left(
- i \, \frac{ \Delta{m}^{2}_{k1} \, L }{ 2 \, p }
\right)
\right|^2
\,,
\label{053}
\end{eqnarray}
where 
$ \Delta{m}^{2}_{k1} \equiv m_k^2 - m_1^2 $
(we take
$ m_1 \leq m_2 \leq \ldots $).
From Eqs.(\ref{052}) and (\ref{053})
it follows that the transition probabilities of
neutrinos and antineutrinos are connected by the relation
\begin{equation}
P_{\nu_\alpha\to\nu_\beta}
=
P_{\bar\nu_\beta\to\bar\nu_\alpha}
\,.
\label{054}
\end{equation}
This relation reflects CPT invariance.

If CP invariance in the lepton sector holds,
then there are conventions
for the arbitrary phases
such that
in the case of Dirac neutrinos
we have
\begin{equation}
U_{{\alpha}k}
=
U_{{\alpha}k}^{*}
\,,
\label{0551}
\end{equation}
whereas
in the case of Majorana neutrinos
we have
\begin{equation}\label{0552}
U_{{\alpha}k}
= -
U_{{\alpha}k}^{*}
\,
\eta_{k}
\,,
\end{equation}
where
$\eta_{k}={\pm}i$
is the CP parity\footnote{
The CP parities of Majorana neutrinos could 
be important for neutrinoless double beta-decay;
for example, if
the $\nu_k$'s
have different CP parities,
their contributions
to the amplitude
of neutrinoless double-beta decay
could cancel each other
\cite{DBB}.}
of the Majorana neutrino with mass $m_k$
(see, for example, Ref. \cite{BP87}).
It is obvious that the CP parities
$\eta_{k}$
do not enter in the expressions
for the transitions amplitudes.
Hence,
in both the Dirac and Majorana cases,
CP invariance
implies that
\cite{BHP80-Doi81}
\begin{equation}
P_{\nu_\alpha\to\nu_\beta}
=
P_{\bar\nu_\alpha\to\bar\nu_\beta}
\,.
\label{056}
\end{equation}

Let us introduce the CP-odd asymmetries
\begin{equation}
D_{\alpha;\beta}\label{Dab}
\equiv
P_{\nu_\alpha\to\nu_\beta}
-
P_{\bar\nu_\alpha\to\bar\nu_\beta}
\,.
\end{equation}
From CPT invariance it follows that
\begin{equation}
D_{\alpha;\beta}
=
-
D_{\beta;\alpha}
\,.
\label{058}
\end{equation}
Furthermore,
from the unitarity of the mixing matrix
we have
\begin{equation}
\sum_{\beta\neq\alpha}
D_{\alpha;\beta}
=
0
\,.
\label{059}
\end{equation}
We observe that in the case of transitions
among three flavour states
($\nu_e$, $\nu_\mu$, $\nu_\tau$)
the CP asymmetries satisfy the relations
\cite{CPrelations} 
\begin{equation}\label{CPrelations}
D_{e;\mu}
=
D_{\mu;\tau}
=
D_{\tau;e}
\,,
\label{060}
\end{equation}
which follow from Eqs.(\ref{058}) and (\ref{059}).

In the general case of mixing
of an arbitrary number of massive neutrinos,
the transition probabilities are given by
\begin{equation}\label{PI}
P_{\stackrel{\scriptscriptstyle(-)}{\nu}_{{\hskip-4pt}\alpha}
\to\stackrel{\scriptscriptstyle(-)}{\nu}_{{\hskip-4pt}\beta}}
=
\sum_j |U_{\alpha j}|^2|U_{\beta j}|^2 +
2 \sum_{k>j} \mbox{Re}\!\left[ U_{{\alpha}j} \, U_{{\beta}j}^{*}
\, U_{{\alpha}k}^{*} \, U_{{\beta}k} \right] 
\cos \frac{ \Delta{m}^{2}_{kj}L }{ 2p }
\pm \frac{1}{2} D_{\alpha;\beta}
\,,
\end{equation}
where the plus (minus) sign
applies to neutrinos (antineutrinos).
The expression for
the asymmetries is
\begin{equation}
D_{\alpha;\beta}\label{061}
= \sum_{k>j} I_{\alpha \beta;jk}
\sin
\frac{ \Delta{m}^{2}_{kj} L }{ 2p }
\,,
\end{equation}
with
\begin{equation}\label{I}
I_{\alpha\beta;jk} \equiv 
4 \, \mbox{Im}\!\left[
U_{{\alpha}j}
\,
U_{{\beta}j}^{*}
\,
U_{{\alpha}k}^{*}
\,
U_{{\beta}k}
\right] \,.
\end{equation}
These parameters are invariant under rephasing
of the neutrino mixing matrix
and
(apart for a factor 4)
are the analogues in the lepton sector of the well-known
rephasing-invariant parameters in the quark sector
\cite{Jarlskog,DGW,Dunietz}.
In Sections
\ref{CP violation in the schemes with four neutrinos}--\ref{Matter effects}
we will derive the constraints on the parameters
$I_{\alpha\beta;jk}$
which follow from the results of neutrino oscillation experiments
in the framework of schemes with four and three massive neutrinos.

CP violation in the lepton sector can be observed
in neutrino oscillation experiments
only if at least one of the terms
of the sum in Eq.(\ref{061})
does not vanish
because of the averaging over the neutrino energy spectrum
and the size of the neutrino source and detector. 

If only one mass-squared difference (denoted by $\Delta{m}^2$)
is relevant for short-baseline neutrino oscillations,
the neutrinos can be divided
in two groups $\nu_1, \ldots , \nu_r$ and 
$\nu_{r+1}, \ldots , \nu_n$ with masses $m_1 \leq \ldots \leq m_r$ and
$m_{r+1} \leq \ldots \leq m_n$,
respectively,
such that in SBL experiments
\begin{equation}
\frac{\Delta{m}^2_{kj}L}{2p} \ll 1 \quad \mbox{for} \quad
j,k \leq r \quad \mbox{or} \quad j,k > r \,,
\end{equation}
whereas
\begin{equation}
\Delta{m}^2_{kj} \simeq \Delta{m}^2 \quad 
\mbox{for} \quad k>r \; \mbox{and} \; j \leq r \,.
\end{equation}
In this case,
for the CP
asymmetries in SBL neutrino oscillation experiments
we obtain \cite{CPrelations}
\begin{equation}
D_{\alpha;\beta}^{(\mathrm{SBL})} = \left(
\sum_{r \geq k>j} + \sum_{k>j>r} + \sum_{k>r \geq j} \right)
I_{\alpha\beta;jk} \sin \frac{\Delta{m}^2_{kj}L}{2p} \simeq
\sum_{k>r \geq j} I_{\alpha\beta;jk} 
\sin \frac{\Delta{m}^2 L}{2p} = 0 \,.
\end{equation}
The last step follows from the definition (\ref{I})
and the unitarity of the mixing matrix.
Consequently,
it is necessary to consider neutrino
oscillations in LBL experiments
in order to have some possibility
to observe effects of CP violation in the lepton sector.

In the rest of this paper we will consider
schemes with four and three massive neutrinos,
in which only the largest mass-squared difference
$\Delta{m}^2$
is relevant for neutrino oscillations in SBL experiments,
having a value
in the wide range
\begin{equation}
10^{-1} \, \mathrm{eV}^2
\leq
\Delta{m}^{2}
\leq
10^{3} \, \mathrm{eV}^2
\,,
\label{widerange}
\end{equation}
which include the interval (\ref{LSNDrange})
allowed by the results of the LSND experiment.

\section{Four massive neutrinos}
\label{Four massive neutrinos}

All existing indications in favour of neutrino
oscillations can be accommodated
by a scheme with mixing of
four massive neutrinos
\cite{four,BGKP,BGG96A,BGG96B,OY96}.
In Refs. \cite{BGG96A,BGG96B}
we have shown that from the six possible spectral schemes
of four massive neutrinos,
which correspond to three
different scales of mass-squared differences
$\Delta{m}^{2}_{kj}$,
only two schemes
are compatible with the results of all experiments
(see also Ref. \cite{OY96}).
In these two schemes
the four neutrino masses
are divided in two pairs of close masses
separated by a gap of
$ \sim 1 \, \mathrm{eV} $:
\begin{equation}\label{spectrum}
\mbox{(A)}
\qquad
\underbrace{
\overbrace{m_1 < m_2}^{\mathrm{atm}}
\ll
\overbrace{m_3 < m_4}^{\mathrm{solar}}
}_{\mathrm{LSND}}
\qquad \mbox{and} \qquad
\mbox{(B)}
\qquad
\underbrace{
\overbrace{m_1 < m_2}^{\mathrm{solar}}
\ll
\overbrace{m_3 < m_4}^{\mathrm{atm}}
}_{\mathrm{LSND}}
\,.
\label{AB}
\end{equation}
In scheme A,
$\Delta{m}^{2}_{21}$
is relevant
for the explanation of the atmospheric neutrino anomaly
and
$\Delta{m}^{2}_{43}$
is relevant
for the suppression of solar $\nu_e$'s.
In scheme B,
the roles of
$\Delta{m}^{2}_{21}$
and
$\Delta{m}^{2}_{43}$
are reversed.

Let us define the quantities
$c_{\alpha}$
(with $\alpha=e,\mu,\tau,s$)
as
\begin{equation}\label{dc}
c_{\alpha}
\equiv
\sum_{k=1,2} |U_{{\alpha}k}|^2
\,.
\end{equation}
Taking into account the results
of SBL neutrino oscillation experiments
and those
of solar and atmospheric neutrino experiments,
in the two schemes A and B
the parameters $c_e$ and $c_\mu$
are constrained by
\cite{BGG96A,BGG96B}
\begin{eqnarray}
\mathrm{(A)} \qquad & c_{e} \leq a^{0}_{e} \,, & 
\qquad c_{\mu} \geq 1 - a^{0}_{\mu} \,, \label{A}\\
\mathrm{(B)} \qquad & \!\qquad c_{e} \geq 1 - a^{0}_{e} \,, & 
\qquad c_{\mu} \leq a^{0}_{\mu} \,, \label{B}
\end{eqnarray}
where
\begin{equation} \label{a0}
a^{0}_{\alpha} = \frac{1}{2}
\left(1-\sqrt{1-B_{\alpha;\alpha}^{0}}\,\right)
\quad (\alpha = e,\mu)
\end{equation}
and $B^0_{\alpha;\alpha}$ is the experimental upper bound
for the oscillation amplitude
\cite{BGG96B} 
\begin{equation}
B_{\alpha;\alpha} = 4\, c_\alpha (1-c_\alpha)
\end{equation}
in SBL disappearance experiments.
The values of
$a^{0}_{e}$
and
$a^{0}_{\mu}$
obtained,
respectively,
from the 90\% exclusion plots of
the
Bugey \cite{Bugey95}
$\bar\nu_e\to\bar\nu_e$
reactor experiment
and of the
CDHS \cite{CDHS84} and CCFR \cite{CCFR84}
$\nu_\mu\to\nu_\mu$
accelerator experiments
are given in Fig. 1 of Ref. \cite{BBGK96}.
From that figure one can see that
$a^{0}_{e}$
is small
($ a^{0}_e \lesssim 4 \times 10^{-2} $)
for
$\Delta{m}^{2}$
in the wide range (\ref{widerange})
and
$a^{0}_{\mu}$
is small
($ a^{0}_\mu \lesssim 10^{-1} $)
for
$
\Delta{m}^{2} \gtrsim 0.5 \, \mathrm{eV}^2
$.
In the following
we will use also the upper bounds $A^0_{\alpha;\beta}$
for $\alpha = \mu$ and $\beta = e, \tau$
on the SBL oscillation amplitudes \cite{BGG96B}
\begin{equation} \label{A12}
A_{\alpha;\beta}
=
4 \left| \sum_{k=1,2} U_{{\beta}k} U_{{\alpha}k}^{*} \right|^2
=
4 \left| \sum_{k=3,4} U_{{\beta}k} U_{{\alpha}k}^{*} \right|^2
\,,
\end{equation}
which are obtained from 90\% CL exclusion plots of the
BNL E734
\cite{BNLE734},
BNL E776
\cite{BNLE776}
and
CCFR
\cite{CCFR96}
$
\stackrel{\makebox[0pt][l]
{$\hskip-3pt\scriptscriptstyle(-)$}}{\nu_{\mu}}
\to\stackrel{\makebox[0pt][l]
{$\hskip-3pt\scriptscriptstyle(-)$}}{\nu_{e}}
$
appearance experiments
and of the
FNAL E531
\cite{FNALE531}
and
CCFR
\cite{CCFR95}
$
\stackrel{\makebox[0pt][l]
{$\hskip-3pt\scriptscriptstyle(-)$}}{\nu_{\mu}}
\to\stackrel{\makebox[0pt][l]
{$\hskip-3pt\scriptscriptstyle(-)$}}{\nu_{\tau}}
$
appearance experiments.
The result of the LSND
\cite{LSND}
$ \bar\nu_\mu \to \bar\nu_e $
appearance experiment,
which is crucial for the arguments in favour
of the 4-neutrino schemes A and B
\cite{BGG96A,BGG96B,OY96}, 
will be of importance in the further discussion
mainly through the allowed range (\ref{LSNDrange}) of
$\Delta{m}^2$.

It can be seen from the neutrino mass spectra (\ref{spectrum})
that the expressions of the oscillation probabilities in scheme B
follow from the corresponding expressions in scheme A
through the exchange of indices
\begin{equation}\label{0721}
1 \, , \, 2
\leftrightarrows
3 \, , \, 4
\,.
\end{equation}
Since this permutation of indices transforms the conditions (\ref{A}) 
into $1-c_e \leq a^0_e$ and $1-c_\mu \geq 1-a^0_\mu$,
which are identical with those of Eq.(\ref{B}),
it follows that the schemes
A and B are indistinguishable with neutrino oscillations
\cite{BGG97}.
Note, however, that these schemes could in principle
be distinguished, for instance, in $(\beta\beta)_{0\nu}$ decay or
with the measurement
of the high-energy part of the $\beta$-spectrum
of $^3\mathrm{H}$ \cite{BGG96A,BGG96B}. In the following
we will perform all calculations in scheme A,
but all bounds on CP-violating observables which we will derive
hold in both schemes.

Short-baseline neutrino oscillation experiments are sensitive to
$\Delta{m}^2 \equiv \Delta{m}^{2}_{41} \gtrsim 0.1 \, \mathrm{eV}^2$
with a distance $L$ between neutrino source and detector such that
\begin{equation} \label{ass}
\frac{ \Delta{m}^{2}_{21} \, L }{ 2 \, p } \ll 1
\qquad
\mbox{and}
\qquad
\frac{ \Delta{m}^{2}_{43} \, L }{ 2 \, p } \ll 1
\,.
\end{equation}
As discussed in the previous section,
with these assumptions on
the neutrino mass spectrum
there are no effects of CP
violation in SBL neutrino oscillations.
On the other hand,
long-baseline neutrino oscillation experiments are
planned to be sensitive
to the ``atmospheric neutrino range''
$
10^{-3} \, \mathrm{eV}^2
\lesssim
\Delta{m}^{2}_{kj}
\lesssim
10^{-1} \, \mathrm{eV}^2
$.
In scheme A,
the probabilities of
$ \nu_\alpha \to \nu_\beta $
and
$ \bar\nu_\alpha \to \bar\nu_\beta $
transitions
in LBL experiments
are given by
\begin{eqnarray}
&&
P^{(\mathrm{LBL,A})}_{\nu_\alpha\to\nu_\beta}
=
\left|
U_{\beta1}
\,
U_{\alpha1}^{*}
+
U_{\beta2}
\,
U_{\alpha2}^{*}
\,
\exp\!\left(
- i
\frac{ \Delta{m}^{2}_{21} \, L }{ 2 \, p }
\right)
\right|^2
+
\left|
\sum_{k=3,4}
U_{{\beta}k}
\,
U_{{\alpha}k}^{*}
\right|^2
\,,
\label{plba1}
\\
&&
P^{(\mathrm{LBL,A})}_{\bar\nu_\alpha\to\bar\nu_\beta}
=
\left|
U_{\beta1}^{*}
\,
U_{\alpha1}
+
U_{\beta2}^{*}
\,
U_{\alpha2}
\,
\exp\!\left(
- i
\frac{ \Delta{m}^{2}_{21} \, L }{ 2 \, p }
\right)
\right|^2
+
\left|
\sum_{k=3,4}
U_{{\beta}k}^{*}
\,
U_{{\alpha}k}
\right|^2
\,,
\label{plba2}
\end{eqnarray}
respectively.
Matter effects are not included in these formulas which
have been obtained from Eqs.(\ref{052}) and (\ref{053}),
respectively, by
taking into account that in LBL experiments
$ \Delta{m}^{2}_{43} L / 2 p \ll 1 $
and
averaging out the oscillating terms with phases much larger
than $2\pi$
($ \Delta{m}^{2}_{kj} L / 2 p \gg 2\pi $
for $k=3,4$ and $j=1,2$).

From Eqs.(\ref{plba1}), (\ref{plba2}) and (\ref{0721})
it follows that the CP-odd asymmetries
$D^{(\mathrm{LBL})}_{\alpha;\beta}$
in LBL experiments
in the schemes A and B are given by
\begin{eqnarray}
&&
D^{(\mathrm{LBL,A})}_{\alpha;\beta}
=
I^{(\mathrm{A})}_{\alpha\beta}
\,
\sin
\frac{ \Delta{m}^{2}_{21}L }{ 2p }
\,,
\label{073}
\\
&&
D^{(\mathrm{LBL,B})}_{\alpha;\beta}
=
I^{(\mathrm{B})}_{\alpha\beta}
\,
\sin
\frac{ \Delta{m}^{2}_{43}L }{ 2p }
\,,
\label{074}
\end{eqnarray}
with
the oscillation amplitudes (see Eq.(\ref{I}))
\begin{equation}\label{iab}
I^{(\mathrm{A})}_{\alpha\beta} \equiv I_{\alpha\beta;12}
\quad \mbox{and} \quad
I^{(\mathrm{B})}_{\alpha\beta} \equiv I_{\alpha\beta;34} \,.
\end{equation}
In the following we will only study scheme A and drop the 
superscript A 
($I^{(\mathrm{A})}_{\alpha\beta} \equiv I_{\alpha\beta}$) 
for the reasons mentioned above in the context of Eq.(\ref{0721}).

Finally, we want to mention that the phases
of the products of elements of the mixing matrix
whose imaginary parts give
the quantities $I_{e\mu}$, $I_{e\tau}$ and
$I_{\mu\tau}$ are not independent \cite{Dunietz},
as can be seen from the obvious relation
\begin{equation}\label{arg}
\arg \left[
U_{\mu 1}\,U_{\tau 1}^*\,U_{\mu 2}^*\,U_{\tau 2}
\right] =
\arg \left[
U_{e 1}\,U_{\tau 1}^*\,U_{e 2}^*\,U_{\tau 2}
\right] -
\arg \left[
U_{e 1}\,U_{\mu 1}^*\,U_{e 2}^*\,U_{\mu 2}
\right]
\,,
\end{equation}
which is valid if $I_{e\mu}$, $I_{e\tau}$ and
$I_{\mu\tau}$ are all different from zero.
Hence, a measurement of
$I_{e\mu}$, $I_{e\tau}$ and $I_{\mu\tau}$
can give information on only two independent linear combinations
of the three CP-violating phases
which are possible in the four-neutrino schemes.
In order to obtain information on the values of all
the three CP-violating phases
it is necessary to measure also some of the other
$I_{\alpha\beta;kj}$'s.

\section{CP violation in the schemes with four neutrinos}
\label{CP violation in the schemes with four neutrinos}

In this section we discuss bounds on the vacuum quantities
(\ref{iab}). For the reasons explained in 
Sect. \ref{Four massive neutrinos} we confine ourselves to
scheme A in the derivations of the bounds.
As shown in Appendix \ref{apb},
the unitarity of the mixing matrix
implies the
``unitarity bound''
\begin{equation}
|I_{\alpha\beta}| \leq f(c_\alpha,c_\beta)
\label{Kfcc}
\end{equation}
where $f(x,y)$ is the continuous function 
\begin{equation}
f(x,y) = 
\left\{ \begin{array}{lcl} \displaystyle 
f_1 \equiv xy
& \quad \mbox{for} \quad &
2(1-x)(1-y) \ge xy
\\[3mm] \displaystyle
f_2 \equiv 2 [(x+y-1)(1-x)(1-y)]^{1/2}
& \quad \mbox{for} \quad &
2(1-x)(1-y) < xy
\end{array} \right.
\label{fxy}
\end{equation}
defined on the unit square $0 \le x \le 1$, $0 \le y \le 1$.
In Fig. \ref{fig1}
we have drawn a contour plot of the function
$f(x,y)$,
which is helpful for the determination of the maximal
allowed value for
$f(c_\alpha , c_\beta)$
when
$c_\alpha$ and/or $c_\beta$
are bounded.
The dotted line
in Fig. \ref{fig1}
is the borderline
$g(x)=2(1-x)/(2-x)$
between the regions where
$f=f_{1}$
and
$f=f_{2}$.
Note that
$f$ is continuous
along this borderline.

In order to determine the maxima of
$f(x,y)$,
the following considerations are useful
(for the details
consult Appendix \ref{apc}).
Increasing $x$ at fixed $y$,
the function $f$ increases monotonously
from $f=0$ at $x=0$,
until the straight line
$y_1(x)=2-2x$ ($ 1/2 \le x \le 1 $) 
depicted in Fig. \ref{fig1} is reached. There, the value of $f$ is
given by $f=y\sqrt{1-y}$.
After this intersection,
the function $f$ decreases monotonously to $f=0$
at $x=1$.
From the symmetry
$f(x,y)=f(y,x)$,
it follows that for fixed $x$ and
increasing $y$ the function $f$ increases monotonously
from $f=0$ at $y=0$ to $f=x\sqrt{1-x}$ when the straight line
$y_2(x)=1-x/2$ ($ 0 \le x \le 1 $)
depicted in Fig. \ref{fig1} is crossed.
After this intersection,
$f$ decreases monotonously to $f=0$
at $y=1$.

The absolute maximum of
the function $f$ (see Appendix \ref{apc})
lies at the intersection
of the lines $y_1$ and $y_2$ and is given by
$
f_{\mathrm{max}}
=
2/3\sqrt{3}
\approx
0.385
$.
Therefore,
from the unitarity of the mixing matrix
we have an absolute maximum for
$|I_{\alpha\beta}|$:
\begin{equation}\label{0002}
|I_{\alpha\beta}|
\leq
\frac{2}{3\sqrt{3}}
\approx
0.385
\,.
\end{equation}

With the help of Fig. \ref{fig1},
one can see that
Eq.(\ref{Kfcc})
with the constraints (\ref{A}) on
$c_{e}$ and $c_{\mu}$
implies that
\begin{equation}\label{Ime1}
|I_{e\mu}|
\leq
\left\{
\begin{array}{lcl} \displaystyle
f_2(a^{0}_{e},y_2(a^{0}_{e})) = 
a^{0}_{e} \left(1-a^{0}_{e}\right)^{1/2}
& \quad \mbox{for} \quad &
a^{0}_{\mu} \geq a^{0}_{e} / 2 
\,,
\\[3mm] \displaystyle
f_2(a^{0}_{e},1-a^{0}_{\mu}) =
2
\left[
\left(a^{0}_{e}-a^{0}_{\mu}\right)
\left(1-a^{0}_{e}\right)
\,
a^{0}_{\mu}
\right]^{1/2}
& \quad \mbox{for} \quad &
a^{0}_{\mu} \leq a^{0}_{e} / 2
\,.
\end{array} \right.
\end{equation}
The solid curve in Fig. \ref{fig2}
shows the limit
$ |I_{e\mu}| \leq a^{0}_{e} \, \sqrt{1-a^{0}_{e}} $
with $a^{0}_{e}$
obtained from
the 90\% CL exclusion plot of the Bugey \cite{Bugey95}
$\bar\nu_{e}\to\bar\nu_{e}$
experiment.
The dash-dotted curve in Fig. \ref{fig2}
represents the
improvement reached with
the lower part of Eq.(\ref{Ime1})
at the values of $\Delta{m}^2$
for which
$ a^{0}_{\mu} \leq a^{0}_{e} / 2 $,
with
$a^{0}_{\mu}$
obtained from the 90\% CL exclusion plots of
the CDHS \cite{CDHS84} and CCFR \cite{CCFR84}
$\nu_\mu\to\nu_\mu$
experiments.
From this figure one can see that
the upper bound for
$|I_{e\mu}|$ is very small
($ |I_{e\mu}| \lesssim 4 \times 10^{-2} $)
for
$\Delta{m}^2$
in the wide range (\ref{widerange}).

The bound represented by the solid curve in Fig. \ref{fig2}
is valid also for
$|I_{e\tau}|$,
because there is no experimental information on $c_{\tau}$.

For $|I_{\mu\tau}|$,
again by inspection of Fig. \ref{fig1},
one can see that
Eq.(\ref{Kfcc})
with the constraint (\ref{A}) on
$c_{\mu}$
implies that
\begin{equation}\label{Imt1}
|I_{\mu\tau}|
\leq
f_2\!\left(1-a^{0}_{\mu}, y_2(1-a^0_\mu) \right) = 
\left(1-a^{0}_{\mu}\right) \sqrt{a^{0}_{\mu}}
\,.
\end{equation}
The solid curve in Fig. \ref{fig3}
represents
the corresponding bound
obtained from the
90\% CL exclusion curves of the
CDHS \cite{CDHS84} and CCFR \cite{CCFR84}
$\nu_\mu\to\nu_\mu$
experiments.
For
$ \Delta{m}^2 \lesssim 0.3 \, \mathrm{eV}^2 $
there are no experimental data
and therefore
$ |I_{\mu\tau}|_{\mathrm{max}} \approx 0.385 $
by virtue of Eq.(\ref{0002}).

Taking into account the expression
(\ref{A12})
for
$A_{\alpha;\beta}$,
in both schemes A and B
we have also the ``amplitude bound''
(for the proof of this inequality,
see Appendix \ref{apa})
\begin{equation}\label{X08}
|I_{\alpha\beta}|
\leq
\frac{ 1 }{ 2 }
\sqrt{
A_{\alpha;\beta}
\left(
4 \, c_{\alpha} \, c_{\beta}
-
A_{\alpha;\beta}
\right)
}
\,.
\end{equation}
With the upper bound
$ A_{\alpha;\beta} \leq A_{\alpha;\beta}^0 $
we obtain
\begin{equation}\label{091}
|I_{\alpha\beta}|
\leq
\left\{
\begin{array}{lcl} \displaystyle
\frac{ 1 }{ 2 }
\sqrt{
A_{\alpha;\beta}^{0}
\left(
4 \, c_{\alpha} \, c_{\beta}
-
A_{\alpha;\beta}^{0}
\right)
}
& \quad \mbox{for} \quad &
A_{\alpha;\beta}^{0}
\leq
2 \, c_{\alpha} \, c_{\beta}
\,,
\\[3mm] \displaystyle
c_{\alpha} \, c_{\beta}
& \quad \mbox{for} \quad &
A_{\alpha;\beta}^{0}
\geq
2 \, c_{\alpha} \, c_{\beta}
\,.
\end{array} \right.
\end{equation}
For
$|I_{e\mu}|$,
with the constraints (\ref{A}),
the inequality (\ref{091})
becomes
\begin{equation}\label{Ime2}
|I_{e\mu}|
\leq
\left\{
\begin{array}{lcl} \displaystyle
\frac{ 1 }{ 2 }
\sqrt{
A_{\mu;e}^{0}
\left(
4 \, a^{0}_{e}
-
A_{\mu;e}^{0}
\right)
}
& \quad \mbox{for} \quad &
A_{\mu;e}^{0}
\leq
2 \, a^{0}_{e}
\,,
\\[3mm] \displaystyle
a^{0}_{e}
& \quad \mbox{for} \quad &
A_{\mu;e}^{0}
\geq
2 \, a^{0}_{e}
\,.
\end{array} \right.
\end{equation}
The dashed curve in Fig. \ref{fig2}
shows the limit (\ref{Ime2})
obtained using the 90\% exclusion plots
of the Bugey \cite{Bugey95}
$\bar\nu_{e}\to\bar\nu_{e}$
experiment for the determination of
$a^{0}_{e}$
and
the
BNL E734
\cite{BNLE734},
BNL E776
\cite{BNLE776}
and
CCFR
\cite{CCFR96}
$\nu_{\mu}\to\nu_{e}$
experiments for the determination of
$A_{\mu;e}^{0}$.
One can see that the upper bound for
$|I_{e\mu}|$ is extremely small
($ |I_{e\mu}| \lesssim 10^{-2} $)
for
$\Delta{m}^2$
in the wide range (\ref{widerange}),
which includes the LSND-allowed range (\ref{LSNDrange}).

Since the constraints (\ref{A})
do not put an upper bound on
the possible values of
$c_{\mu}$
and
$c_{\tau}$,
in the case of
$|I_{\mu\tau}|$
the inequality (\ref{091})
becomes
\begin{equation}\label{Imt2}
|I_{\mu\tau}|
\leq
\frac{ 1 }{ 2 }
\sqrt{
A_{\mu;\tau}^{0}
\left(
4
-
A_{\mu;\tau}^{0}
\right)
}
\,.
\end{equation}
The dashed curve in Fig. \ref{fig3}
shows the limit (\ref{Imt2})
obtained using the 90\% exclusion plots
of the
FNAL E531
\cite{FNALE531}
and
CCFR
\cite{CCFR95}
$\nu_{\mu}\to\nu_{\tau}$
experiments for the determination of
$A_{\mu;\tau}^{0}$.

The shadowed regions in Figs. \ref{fig2} and \ref{fig3}
correspond to the range (\ref{LSNDrange}) of $\Delta{m}^2$
allowed at 90\% CL by the results of the LSND
and all the other SBL experiments.
From Fig. \ref{fig3}
it can be seen that,
taking into account the LSND signal,
$|I_{\mu\tau}|$
could be close to
the maximal value
$ 2/3\sqrt{3} $
allowed
by the unitarity of the mixing matrix.

\section{Three massive neutrinos}
\label{Three massive neutrinos}

Though not all present indications in favour of neutrino mixing
can be taken into account in scenarios with mixing of three massive
neutrinos,
it is nevertheless interesting to investigate also
this case with the methods developed in this paper.
Let us assume for definiteness 
that of the two differences
of squares of neutrino masses one is
relevant for SBL
oscillations
and the other one for LBL
oscillations
(see also Refs. \cite{TA96,AR96,MN97,BGG97}).
These assumptions give rise to the following
two three-neutrino mass spectra
\begin{equation}
(\mbox{\textrm{I}})
\qquad
\underbrace{
\overbrace{m_1 < m_2}^{\mathrm{LBL}}
\ll
m_3
}_{\mathrm{SBL}}
\qquad \mbox{and} \qquad 
(\mbox{\textrm{II}})
\qquad
\underbrace{
m_1
\ll
\overbrace{m_2 < m_3}^{\mathrm{LBL}}
}_{\mathrm{SBL}}
\,.
\label{3nu}
\end{equation}
Furthermore, the results
of the disappearance experiments allow to define the three regions
(see Refs. \cite{BBGK95,BBGK96,BGKP})
\begin{equation}\label{regions}
\begin{array}{lll} \displaystyle
(1) \qquad &
|U_{ek}|^2 \geq 1-a^0_e
\,,
\qquad
&
|U_{{\mu}k}|^2 \leq a^0_\mu
\,,
\\[3mm] \displaystyle
(2) \qquad &
|U_{ek}|^2 \leq a^0_e
\,,
\qquad
&
|U_{\mu k}|^2 \leq a^0_\mu
\,,
\\[3mm] \displaystyle
(3) \qquad &
|U_{ek}|^2 \leq a^0_e
\,,
\qquad
&
|U_{\mu k}|^2 \geq 1-a^0_\mu
\,,
\end{array} 
\end{equation}
with
$k=3$ for the scheme \textrm{I}
and
$k=1$ for the scheme \textrm{II}
(for the definition of $a^0_e$ and $a^0_\mu$,
see Eq.(\ref{a0})).
The neutrino and antineutrino
LBL oscillation probabilities
in scheme \textrm{I}
are given by
\begin{eqnarray}
&&
P^{(\mathrm{LBL},\mathrm{I})}_{\nu_\alpha\to\nu_\beta}
=
\left|
U_{\beta1}
\,
U_{\alpha1}^{*}
+
U_{\beta2}
\,
U_{\alpha2}^{*}
\,
\exp\!\left(
- i
\frac{ \Delta{m}^{2}_{21} \, L }{ 2 \, p }
\right)
\right|^2
+
|U_{{\beta}3}|^2
\,
|U_{{\alpha}3}|^2
\,,
\label{0171}
\\
&&
P^{(\mathrm{LBL},\mathrm{I})}_{\bar\nu_\alpha\to\bar\nu_\beta}
=
\left|
U_{\beta1}^{*}
\,
U_{\alpha1}
+
U_{\beta2}^{*}
\,
U_{\alpha2}
\,
\exp\!\left(
- i
\frac{ \Delta{m}^{2}_{21} \, L }{ 2 \, p }
\right)
\right|^2
+
|U_{{\beta}3}|^2
\,
|U_{{\alpha}3}|^2
\,.
\label{0172}
\end{eqnarray}
From a comparison of Eqs.(\ref{0171}) and (\ref{0172})
with Eqs.(\ref{plba1}) and (\ref{plba2}),
it is obvious that the CP-odd
asymmetries
$D^{(\mathrm{LBL},\mathrm{I})}_{\alpha;\beta}$
are given by the same formulas
(\ref{073}) and (\ref{iab}) as in the 4-neutrino case
with superscript I instead of A.
The transition probabilities in scheme \textrm{II}
are obtained from the expressions
(\ref{0171}) and (\ref{0172})
by a cyclic permutation of the indices:
$1 \to 2 \to 3 \to 1$.
Therefore,
as in the case of the schemes A and B for four neutrinos,
the bounds on the CP-odd parameters
$I_{\alpha\beta}$
are the same in the three neutrino schemes
\textrm{I} and \textrm{II}. 

The methods
for the derivation of the bounds on
$I_{\alpha\beta}$
which are described in the appendices
for the four-neutrino schemes (\ref{AB})
are valid also in the case of mixing of three neutrinos.
Obviously, the derivations
of the four-neutrino case A (B)
are carried over to the three-neutrino case
\textrm{I} (\textrm{II})
if we put
$U_{\alpha4}=0$
($U_{\alpha1}=0$
and change the indices
$2,3,4\to1,2,3$)
for all
$\alpha=e,\mu,\tau$.
This implies that the amplitude bound (\ref{oab})
applies also to the three neutrino schemes (\ref{3nu}).
In order to derive the unitarity bound from Eq.(\ref{oab}),
one must notice that
from the unitarity of the
$3\times 3$ mixing matrix we have
$ A_{\alpha;\beta} = 4(1-c_\alpha)(1-c_\beta) $
and
$ c_\alpha + c_\beta \geq 1 $.
Hence,
the equality sign applies in Eq.(\ref{cs})
and there is no such distinction as defined by Eq.(\ref{ineq}).
Consequently,
by simple substitution of
$ A_{\alpha;\beta} = 4(1-c_\alpha)(1-c_\beta) $
in Eq.(\ref{oab}),
in the case of mixing of three neutrinos
one obtains the unitarity bound (\ref{B6}),
i.e.
\begin{equation}
|I_{\alpha\beta}|
\leq
f_2(c_\alpha,c_\beta)
\,.
\label{u3}
\end{equation}
Notice that
$f_2(c_\alpha,c_\beta)$
vanishes on the unitarity boundary
$ c_\alpha + c_\beta = 1 $.
Since the maxima of
$f(c_\alpha,c_\beta)$ and $f_2(c_\alpha,c_\beta)$
coincide
and are reached for
$ c_\alpha = c_\beta = 2/3 $
(see Appendix \ref{apc}),
the absolute maximum
$ |I_{\alpha \beta}|_\mathrm{max} = 2/3\sqrt{3} $ 
of the 4-neutrino case extends its validity to three
neutrinos\footnote{This value
corresponds to the maximal value of the
Jarlskog parameter 
$J$ \cite{Jarlskog,DGW} for CP violation in the Kobayashi--Maskawa matrix,
$|J|_{\mathrm{max}}=1/6\sqrt{3}$,
with an additional factor 4 due to the definition
(\ref{I}).}.

Furthermore,
since in the 3-neutrino case the CP-odd asymmetries
in different oscillation channels
are connected by Eq.(\ref{CPrelations}), we have
\begin{equation}
I_{e \mu} = I_{\mu \tau} = I_{\tau e}
\,.
\label{III}
\end{equation}

In the following we will give LBL bounds for each of the
regions (\ref{regions}),
along the lines of the previous 4-neutrino section.

\emph{Region 1.}
With respect to SBL and LBL neutrino
oscillations,
the 3-neutrino schemes \textrm{I} and \textrm{II}
in Region 1
correspond to the
4-neutrino schemes A and B, respectively,
with the same bounds on
$|I_{e\mu}|$
(Eqs.(\ref{Ime1}), (\ref{Ime2}) and Fig. \ref{fig2}).
Because of Eq.(\ref{III})
the stringent bounds on $|I_{e\mu}|$
given in Fig. \ref{fig2}
are valid also for $|I_{\mu\tau}|$.

\emph{Region 2.}
Actually, this Region is disfavoured by the results of the
LSND experiment (see Refs. \cite{BBGK95,BBGK96,BGKP,BGG96B}).
Also the results of the atmospheric neutrino experiments
\cite{Kam-atm,IMB,Soudan,SK-atm}
taken together with the results of the CHOOZ experiment
\cite{CHOOZ97}
indicate that this Region is disfavoured.
Indeed,
in this region $c_\tau$ is small
($ c_\tau \leq a_e^0 + a_\mu^0 $)
and the atmospheric neutrino anomaly
can be explained only with dominant
$ \nu_\mu \leftrightarrows \nu_e $
oscillations,
which are forbidden by the
results of the CHOOZ experiment.
Since these evidences could disappear
when the results of future more accurate experiments
will be available,
let us discuss the bounds on CP violation with the
methods described in the appendices.
They are given by the
amplitude bound
\begin{equation}
|I_{e\mu}|
\leq
\frac{1}{2}
\,
\sqrt{ A_{\mu;e}^{0} \left( 4 - A_{\mu;e}^{0} \right) }
\end{equation}
and the unitarity bound
\begin{equation}
|I_{e\mu}|
\leq
f_2\!\left(1-a^{0}_{e},1-a^0_\mu\right)
=
2
\,
\sqrt{ a^0_e \, a^0_\mu \left( 1 - a^0_e - a^0_\mu \right) }
\,.
\end{equation}
Both bounds are less restrictive than
the corresponding bounds in Region 1.
By Eq.(\ref{III}) these bounds hold in
all neutrino transition channels.

\emph{Region 3.}
In this region,
where $c_e \geq 1-a^0_e$ and
$c_\mu \leq a^0_\mu$,
the full set of atmospheric neutrino data cannot be explained
in the framework discussed here \cite{BGG96B}.
Applying nevertheless our methods for obtaining bounds on CP violation,
we get the results
\begin{equation}
|I_{e\mu}|
\leq
\frac{1}{2}
\,
\sqrt{ A_{\mu;e}^{0} \left( 4 \, a^0_\mu - A_{\mu;e}^{0} \right) }
\label{I31}
\end{equation}
and
\begin{equation}
| I_{e \mu}| \leq \left\{
\begin{array}{lcl} \displaystyle
a^0_\mu
\,
\left( 1 - a^0_\mu \right)^{1/2}
& \quad \mbox{for} \quad & 
a^0_e \geq a^0_\mu / 2
\,,
\\[1mm] \displaystyle
2
\,
\left[ \left( a^0_\mu - a^0_e \right)
\left( 1 - a^0_\mu \right) a^0_e \right]^{1/2}
& \quad \mbox{for} \quad & 
a^0_e \leq a^0_\mu / 2
\,.
\end{array} \right.
\label{I32}
\end{equation}
The amplitude bound is more stringent than the one in Region 2,
but less restrictive than the one in Region 1
and in the 4-neutrino schemes A and B
(for $ a_e^0 < a_\mu^0 $).
From Eq.(\ref{III}) it follows that
the bounds (\ref{I31}) and (\ref{I32})
are valid also for the parameter
$|I_{\mu\tau}|$
that characterizes the CP-odd asymmetry in the
$\nu_\mu\to\nu_\tau$
channel.

Summarizing the results of this section, we have shown that in
regions 2 and 3 the strong bounds of region 1
and of the 4-neutrino schemes A and B on $|I_{e\mu}|$
are somewhat relaxed. However, regions 2 and 3 are disfavoured
by present hints for neutrino oscillations.

Let us also emphasize that
the solar neutrino problem cannot be explained
by neutrino oscillations
in the three-neutrino schemes considered in this section.
Hence,
we regard them as remote possibilities.

\section{Matter effects and CP violation}
\label{Matter effects}

Since in LBL neutrino oscillation experiments the neutrino beam
travels a long distance through the earth's crust, matter effects
influence the neutrino oscillation probabilities.
The effective Hamiltonians in the flavour basis of 
neutrinos and antineutrinos in the case of mixing of 
four neutrinos are given, respectively,
by\footnote{Since
active and sterile neutrinos are present
in the schemes under consideration, 
both charged-current and neutral-current interactions
contribute to the effective Hamiltonians.}
\begin{eqnarray}
&&
H_{\nu} = \frac{1}{2p} \left( U \hat{M}^2 U^\dagger + 
\mathrm{diag}\, (a_{CC}, 0, 0, a_{NC}) \right)
\,,
\label{Hnu}
\\
&&
H_{\bar\nu} = \frac{1}{2p} \left( U^* \hat{M}^2 U^T - 
\mathrm{diag}\, (a_{CC}, 0, 0, a_{NC}) \right)
\,, 
\label{Hanu}
\end{eqnarray}
where $a_{CC}$
and $a_{NC}$ are given by
\cite{MSW}
\begin{eqnarray}
&&
a_{CC}
=
2\sqrt{2} \, G_F \, N_e \, p
\simeq
2.3 \times 10^{-4} \, \mathrm{eV}^2
\left( \frac{ \rho }{ 3 \, \mathrm{g} \, \mathrm{cm}^{-3} } \right)
\left( \frac{ p }{ 1 \, \mathrm{GeV} } \right)
\,,
\label{acc}
\\
&&
a_{NC}
=
\sqrt{2} \, G_F \, N_n \, p
\simeq
\frac{1}{2} \, a_{CC} \,,
\label{anc}
\end{eqnarray}
respectively, and
$ \hat{M}^2 = \mathrm{diag}(m_1^2,m_2^2,m_3^2,m_4^2) $.
Here
$G_F$ is the Fermi constant,
$N_e$ and $N_n$ are the electron and neutron number
density,
respectively,
and
$\rho$ is the density of matter.
With an average density of
approximately $3\, \textrm{g}\,\textrm{cm}^{-3}$ 
in the lithosphere we get
\begin{equation}\label{matterphase}
\frac{a_{CC}L}{2p} \simeq 0.58 \times 10^{-3} 
\left( \frac{L}{1\, \mathrm{km}} \right) \,.
\end{equation}
Therefore, large matter effects are to be expected for baselines
$L \gtrsim 1000$ km.

In the following
we apply the simplifying approximation
of constant electron and neutron number densities,
which is rather accurate in the case of LBL experiments.
In order to obtain the neutrino oscillation probabilities in matter
we have to diagonalize the Hamiltonians (\ref{Hnu}) and (\ref{Hanu}) 
with unitary matrices $U'$ and $\bar{U}'$, respectively,
leading to 
the eigenvalues
$\epsilon_j/2p$ of $H_\nu$
and
$\bar\epsilon_j/2p$ of $H_{\bar\nu}$.
Thus, we have
\begin{equation}
H_\nu = U' \frac{\hat{\epsilon}}{2p} U'^\dagger
\quad \mbox{and} \quad
H_{\bar\nu} = \bar{U}'^{*} \frac{\hat{\bar\epsilon}}{2p} \bar{U}'^{T} 
\end{equation}
with the diagonal matrices 
$
\hat{\epsilon}
=
\mathrm{diag}
(\epsilon_1,\epsilon_2,\epsilon_3,\epsilon_4)
$ 
and
$
\hat{\bar\epsilon}
=
\mathrm{diag}
(\bar{\epsilon}_1,\bar{\epsilon}_2,\bar{\epsilon}_3,\bar{\epsilon}_4)
$
for neutrinos and antineutrinos, respectively
(in the limit
$ a_{CC} , a_{NC} \to 0 $
of vanishing matter effects we get
$ U' , \bar{U}' \to U $
and
$ \epsilon_j , \bar\epsilon_j \to m^2_j $).

Analogously to the definition of $I_{\alpha\beta;jk}$
in Eq.(\ref{I}),
it is useful to define
\begin{equation} \label{I'}
I'_{\alpha\beta;jk} \equiv 4\, \mbox{Im} \! \left[ 
U'_{\alpha j}\,U'^*_{\beta j}\,U'^*_{\alpha k}\,U'_{\beta k} \right]
\quad \mbox{and} \quad
\bar{I}'_{\alpha\beta;jk} \equiv 4\, \mbox{Im} \! \left[ 
\bar{U}'_{\alpha j}\,\bar{U}'^*_{\beta j}\,
\bar{U}'^*_{\alpha k}\,\bar{U}'_{\beta k} \right]\,.
\end{equation}
The usefulness of these definitions lies in the fact that
\emph{for the proof of
the existence of CP violation due to neutrino mixing
the quantities
$I'_{\alpha\beta;jk}$
and
$\bar{I}'_{\alpha\beta;jk}$
are as good as $I_{\alpha\beta;jk}$}.
In other words,
$I_{\alpha\beta;jk}=0$
for all values of the indices $\alpha,\beta,j,k$
if and only if
$I'_{\alpha'\beta';j'k'}=0$
($\bar{I}'_{\alpha'\beta';j'k'}=0$)
for all values of $\alpha',\beta',j',k'$.
Let us prove this statement
for
$I'_{\alpha\beta;jk}$
(an analogous proof holds in the case of
$\bar{I}'_{\alpha\beta;jk}$).
First we assume that
$I_{\alpha\beta;jk}=0$
for all values of $\alpha,\beta,j,k$.
This is possible only if the mixing matrix $U$
can be written as
$ U = e^{i\rho} \tilde{U} e^{i\sigma} $,
where $\rho$ and $\sigma$
are real diagonal matrices and $\tilde{U}$ is real.
Since
the matter part in the Hamiltonian (\ref{Hnu}) is real and diagonal,
the effective Hamiltonian (\ref{Hnu}) can be written as
\begin{equation}\label{Ht}
H_\nu = e^{i\rho} \frac{1}{2p} \left(
\tilde{U} \hat{M}^2 \tilde{U}^\dagger + 
\mathrm{diag} (a_{CC}, 0, 0, a_{NC}) \right)
e^{-i\rho} \equiv e^{i\rho} \tilde{H}_\nu e^{-i\rho}
\,,
\end{equation}
where $\tilde{H}_\nu$ is a real and symmetric matrix.
Hence,
the matrix $U'$
can be written as
$U'=e^{i\rho}\tilde{U}'$ where $\tilde{U}'$ is real.
This form of $U'$ implies that
$I'_{\alpha\beta;jk} = 0$
for all values of $\alpha,\beta,j,k$.
In order to prove the inverse statement,
i.e.
that
$I_{\alpha\beta;jk}=0$
for all values of $\alpha,\beta,j,k$
if
$I'_{\alpha'\beta';j'k'}=0$
for all values of $\alpha',\beta',j',k'$,
we note that
\begin{equation}
U \hat{M}^2 U^\dagger = U' \hat{\epsilon} U'^\dagger -
\mathrm{diag}(a_{CC}, 0, 0, a_{NC}) \, .
\end{equation}
Now the r\^{o}les of $U$ and $U'$ are exchanged.
Hence,
the same
reasoning as before leads to the inverse statement.

The possibility of finding evidence for
CP violation in the lepton sector
through the quantities
$I'_{\alpha\beta;jk}$
and
$\bar{I}'_{\alpha\beta;jk}$
leads us to the search for
methods that could allow to extract these quantities from
the transition probabilities
of neutrinos and antineutrinos
measured
in long-baseline oscillation experiments.

The formula in Eq.(\ref{PI}) for the probability of
$
\stackrel{\scriptscriptstyle(-)}{\nu}_{{\hskip-4pt}\alpha}
\to
\stackrel{\scriptscriptstyle(-)}{\nu}_{{\hskip-4pt}\beta}
$
transitions
is adapted to the matter case by the substitutions
$U \to U'$ and $I \to I'$ for neutrinos and
$U \to \bar{U}'$ and $I \to \bar{I}'$ for antineutrinos:
\begin{eqnarray}
&&
P_{\nu_\alpha\to\nu_\beta}
=
\sum_j |U'_{\alpha j}|^2 |U'_{\beta j}|^2
+
2
\sum_{k>j}
\mbox{Re}\big[
U'_{{\alpha}j}
U'^{*}_{{\beta}j}
U'^{*}_{{\alpha}k}
U'_{{\beta}k}
\big] 
\cos \frac{ \epsilon_{kj} L }{ 2p }
+
\frac{1}{2}
\sum_{k>j}
I'_{\alpha\beta;jk}
\sin \frac{ \epsilon_{kj} L }{ 2p }
\,,
\label{P1}
\\
&&
P_{\bar\nu_\alpha\to\bar\nu_\beta}
=
\sum_j |\bar{U}'_{\alpha j}|^2 |\bar{U}'_{\beta j}|^2
+
2
\sum_{k>j}
\mbox{Re}\big[
\bar{U}'_{{\alpha}j}
\bar{U}'^{*}_{{\beta}j}
\bar{U}'^{*}_{{\alpha}k}
\bar{U}'_{{\beta}k}
\big] 
\cos \frac{ \bar\epsilon_{kj} L }{ 2p }
-
\frac{1}{2}
\sum_{k>j}
\bar{I}'_{\alpha\beta;jk}
\sin \frac{ \bar\epsilon_{kj} L }{ 2p }
\,,
\label{P2}
\end{eqnarray}
with
$ \epsilon_{kj} \equiv \epsilon_k - \epsilon_j $
and
$ \bar\epsilon_{kj} \equiv \bar\epsilon_k - \bar\epsilon_j $.
From these two equations it is clear that
in matter
the transition probabilities of neutrinos
and antineutrinos are different
even if CP is conserved,
i.e. if all the quantities
$I'_{\alpha\beta;jk}$
and
$\bar{I}'_{\alpha\beta;jk}$
are equal to zero.
Hence,
simple measurements
of the asymmetries (\ref{Dab})
do not allow to obtain direct information
on CP violation
\cite{TA96,AR96,MN97}.
This is due to the fact that the matter contribution
to the effective neutrino and antineutrino
Hamiltonians (\ref{Hnu}), (\ref{Hanu})
is not CP-symmetric.
However,
since
the matter contribution to the effective neutrino (antineutrino)
Hamiltonian is real
and
the matter density is symmetric along the path
of the neutrino beam in terrestrial
long-baseline experiments,
matter effects are T-symmetric \cite{KP87}.
In other words,
there is no difference in the matter contributions to the
$
\stackrel{\scriptscriptstyle(-)}{\nu}_{{\hskip-4pt}\alpha} 
\to
\stackrel{\scriptscriptstyle(-)}{\nu}_{{\hskip-4pt}\beta}
$
and
$
\stackrel{\scriptscriptstyle(-)}{\nu}_{{\hskip-4pt}\beta} 
\to
\stackrel{\scriptscriptstyle(-)}{\nu}_{{\hskip-4pt}\alpha}
$
channels
and a difference of the corresponding transition probabilities
can only be due to a fundamental violation of T in the lepton sector.
Since the CPT theorem implies that a violation of T
is equivalent to a violation of CP,
we are lead to explore the possibility
to obtain direct evidence of CP and T violation in the lepton sector
through measurements of the T-odd asymmetries
\begin{equation}
T_{\alpha\beta}
\equiv
P_{\nu_\alpha\to\nu_\beta} - P_{\nu_\beta\to\nu_\alpha}
\quad \mbox{and} \quad
\bar{T}_{\alpha\beta}
\equiv
P_{\bar\nu_\alpha\to\bar\nu_\beta} - P_{\bar\nu_\beta\to\bar\nu_\alpha}
\label{Todd}
\end{equation}
in long-baseline oscillation experiments.
In the case of a constant matter density along the neutrino path
(which is a good approximation for
terrestrial LBL experiments with a baseline shorter than
about 4000 km),
from Eqs.(\ref{P1}) and (\ref{P2})
it is straightforward to obtain the following expressions
for the T-odd asymmetries:
\begin{equation}
T_{\alpha\beta}
=
\sum_{k>j}
I'_{\alpha\beta;jk}
\,
\sin \frac{ \epsilon_{kj} L }{ 2p }
\quad \mbox{and} \quad
\bar{T}_{\alpha\beta}
=
\sum_{k>j}
\bar{I}'_{\alpha\beta;jk}
\,
\sin \frac{ \bar\epsilon_{kj} L }{ 2p }
\,.
\label{Toddc}
\end{equation}
It is clear that finding
$T_{\alpha\beta}$ and/or $\bar{T}_{\alpha\beta}$
different from zero would be a direct evidence
of T (and CP) violation in the lepton sector
independent from matter effects.

The T-odd asymmetries (\ref{Todd})
cannot be measured in the
accelerator LBL experiments
of the first generation
\cite{K2K,MINOS,ICARUS}
because the initial beam will contain almost exclusively
$\nu_\mu$.
In order to have some possibility to measure
the T-odd asymmetries
it will be necessary to wait for the second generation
of accelerator LBL experiments,
as those that will use
$ \nu_\mu + \bar\nu_e $
and
$ \bar\nu_\mu + \nu_e $
neutrino beams
from a muon collider
\cite{Geer,Mohapatra97}.

Let us now consider
the four-neutrino schemes (\ref{AB}),
whose phenomenology of CP violation
in long-baseline neutrino oscillation experiments
in vacuum
has been discussed in Sections
\ref{Four massive neutrinos}
and
\ref{CP violation in the schemes with four neutrinos}.
We will discuss explicitly only
the neutrino T-odd asymmetries
$T_{\alpha\beta}$
in the scheme A,
but all the conclusions are valid also for
the antineutrino T-odd asymmetries
$\bar{T}_{\alpha\beta}$
in scheme A
and for the neutrino and antineutrino
T-odd asymmetries in scheme B
(with the exchange of indices given in Eq.(\ref{0721})).

In Ref. \cite{BGG97}
we have shown
that apart from small corrections of order $a_{CC}/\Delta{m}^2$,
one can decompose $U'$ into
\begin{equation}\label{R}
U' = UR \quad \mbox{with} \quad
R = \left( \begin{array}{cc}    
R_{\mathrm{atm}} & 0 \\ 0 & R_{\mathrm{sun}} \end{array} \right),
\end{equation}
where $R_{\mathrm{atm}}$ and $R_{\mathrm{sun}}$ are 
2$\times$2 unitary matrices.
The block structure of $R$ implies that
\begin{equation}\label{scalarp}
\sum_{j=1,2} U'_{\alpha j}U'^*_{\beta j} =
\sum_{j=1,2}  U_{\alpha j} U^*_{\beta j}
\end{equation}
and therefore
\begin{equation}\label{ca}
c_\alpha \equiv \sum_{j=1,2} |U_{\alpha j}|^2 = 
\sum_{j=1,2} |U'_{\alpha j}|^2 \, .
\end{equation}
Since the bounds on
$I_{\alpha\beta} \equiv I_{\alpha\beta;12}$
derived in Section
\ref{CP violation in the schemes with four neutrinos}
depend only on quantities of the type (\ref{scalarp}) and (\ref{ca}),
we arrive
at the interesting conclusion that
the upper bounds on
$I_{\alpha\beta}$ are also valid for
\begin{equation}
I'_{\alpha\beta} \equiv I'_{\alpha\beta;12}
\quad \mbox{and} \quad
\bar{I}'_{\alpha\beta} \equiv \bar{I}'_{\alpha\beta;12}
\,.
\label{I'ab}
\end{equation}
In this sense the upper bounds
represented by the curves in Figs.
\ref{fig2} and \ref{fig3} include matter effects.

Using a method analogous to that employed for the derivation of
Eq.(\ref{plba1}),
for the probability of
$\nu_\alpha\to\nu_\beta$
LBL transitions in matter
we obtain
\begin{eqnarray}
P^{(\mathrm{LBL})}_{\nu_\alpha\to\nu_\beta} & = &
\left| U'_{\beta 1}U'^*_{\alpha 1} + U'_{\beta 2}U'^*_{\alpha 2} 
e^{-i\phi} \right|^2 + 
\left| U'_{\beta 3}U'^*_{\alpha 3} + U'_{\beta 4}U'^*_{\alpha 4} 
e^{-i\omega} \right|^2 \nonumber \\
& = &
\sum_k |U'_{\beta k}|^2 |U'_{\alpha k}|^2
+ 2 \, \mathrm{Re}
[U'_{\alpha 1}U'^*_{\beta 1}U'^*_{\alpha 2}U'_{\beta 2}]
\cos \phi
+ 2 \, \mathrm{Re}
[U'_{\alpha 3}U'^*_{\beta 3}U'^*_{\alpha 4}U'_{\beta 4}]
\cos \omega
\nonumber \\
&&
+
\frac{1}{2} \, I'_{\alpha\beta} \sin \phi
+
\frac{1}{2} \, J'_{\alpha\beta} \sin \omega
\,,
\label{plbam}
\end{eqnarray}
with the definitions
\begin{equation}\label{po}
\phi \equiv \frac{\epsilon_2 - \epsilon_1}{2p}L
\,, \quad
\omega \equiv \frac{\epsilon_4 - \epsilon_3}{2p}L
\quad \mathrm{and} \quad
J'_{\alpha\beta}
\equiv
I'_{\alpha\beta;34}
\,.
\end{equation}
The expressions for the quantities
$ \epsilon_2 - \epsilon_1 $
and
$ \epsilon_4 - \epsilon_3 $
in terms of the neutrino mixing parameters
and of the matter density
have been derived in Ref. \cite{BGG97}.

From Eq.(\ref{plbam}),
for the neutrino T-odd asymmetries we obtain the expression
\begin{equation}
T_{\alpha\beta}
=
I'_{\alpha\beta} \sin \phi
+
J'_{\alpha\beta} \sin \omega
\,.
\label{Tnu}
\end{equation}
Therefore,
in matter the T-odd asymmetry
$T_{\alpha\beta}$
depends not only on the parameter
$I'_{\alpha\beta}$
relative to the atmospheric sector
(see Eq.(\ref{R}) and the definitions (\ref{I'}) and (\ref{I'ab})),
but also on the parameter
$J'_{\alpha\beta}$
relative to the solar sector.
In Ref. \cite{BGG97} we have shown that
in the case of accelerator LBL experiments
the maximal value of the parameter $\omega$
is given by
\begin{equation}
\omega_{\mathrm{max}} \simeq \frac{3}{2} \, \frac{a_{CC}L}{2p}
=
\frac{3}{2}
\sqrt{2} \, G_F \, N_e \, L
=
8.6 \times 10^{-4}
\left( \frac{ L }{ 1 \mathrm{km} } \right)
\label{omax}
\end{equation}
for
$ \rho = 3 \, \mathrm{g} \, \mathrm{cm}^{-3} $.
Notice that the value of
$\omega_{\mathrm{max}}$
does not depend on the neutrino energy,
but only on the propagation distance $L$.
From Eq.(\ref{omax}) one can see that
the contribution of the term in Eq.(\ref{Tnu})
proportional to $\sin\omega$
could be relevant for
$ L \gtrsim 100 \, \mathrm{km} $
and therefore cannot be neglected in
the analysis of the results of LBL experiments.

As we have seen above,
the upper bounds on
$ I_{\alpha\beta} \equiv I_{\alpha\beta;12} $
are also valid for
$I'_{\alpha\beta}$
and
$\bar{I}'_{\alpha\beta}$.
An analogous reasoning
leads to the conclusion that
the upper bounds on the parameters
\begin{equation}
J_{\alpha\beta} \equiv I_{\alpha\beta;34}
\label{Jab}
\end{equation}
are also valid for
$J'_{\alpha\beta}$
and
$\bar{J}'_{\alpha\beta} \equiv \bar{I}'_{\alpha\beta;34}$.
Hence we are lead to the investigation of the upper bounds
for the parameters
$J_{\alpha\beta}$.
Following the methods presented in the
Appendices \ref{apa}--\ref{apc},
one can derive the unitarity bound
\begin{equation}
|J_{\alpha\beta}|
\leq
f(1-c_\alpha,1-c_\beta)
\label{u34}
\end{equation}
and the amplitude bound
\begin{equation}
|J_{\alpha\beta}|
\leq
\frac{1}{2} \,
\sqrt{
A_{\alpha;\beta} 
\left[
4 \, (1-c_{\alpha}) \, (1-c_{\beta}) - A_{\alpha;\beta}
\right]
}
\,.
\label{a34}
\end{equation}

Taking into account the constraints (\ref{A})
on $c_e$ and $c_\mu$
obtained from the exclusion curves of SBL reactor and accelerator
disappearance experiments,
the unitarity bound for
$|J_{e\mu}|$
is given by
\begin{equation}\label{Jme1}
|J_{e\mu}|
\leq
\left\{
\begin{array}{lcl} \displaystyle
f_2(a^{0}_{\mu},y_2(a^{0}_{\mu})) = 
a^{0}_{\mu} \left(1-a^{0}_{\mu}\right)^{1/2}
& \quad \mbox{for} \quad &
a^{0}_{e} \geq a^{0}_{\mu} / 2 
\,,
\\[3mm] \displaystyle
f_2(a^{0}_{\mu},1-a^{0}_{e}) =
2
\left[
\left(a^{0}_{\mu}-a^{0}_{e}\right)
\left(1-a^{0}_{\mu}\right)
\,
a^{0}_{e}
\right]^{1/2}
& \quad \mbox{for} \quad &
a^{0}_{e} \leq a^{0}_{\mu} / 2
\,.
\end{array} \right.
\end{equation}
The numerical value of this bound,
obtained from
the 90\% CL exclusion plot of the Bugey \cite{Bugey95}
$\bar\nu_{e}\to\bar\nu_{e}$
experiment
and from
the 90\% CL exclusion plots of the
CDHS \cite{CDHS84} and CCFR \cite{CCFR84}
$\nu_\mu\to\nu_\mu$
experiments,
is shown by the solid curve in Fig. \ref{fig4}.
The dashed curve in Fig. \ref{fig4}
represents the amplitude bound
\begin{equation}\label{Jme2}
|J_{e\mu}|
\leq
\left\{
\begin{array}{lcl} \displaystyle
\frac{ 1 }{ 2 }
\sqrt{
A_{\mu;e}^{0}
\left(
4 \, a^{0}_{\mu}
-
A_{\mu;e}^{0}
\right)
}
& \quad \mbox{for} \quad &
A_{\mu;e}^{0}
\leq
2 \, a^{0}_{\mu}
\,,
\\[3mm] \displaystyle
a^{0}_{\mu}
& \quad \mbox{for} \quad &
A_{\mu;e}^{0}
\geq
2 \, a^{0}_{\mu}
\,,
\end{array} \right.
\end{equation}
obtained from
the 90\% CL exclusion plots of the
CDHS \cite{CDHS84} and CCFR \cite{CCFR84}
$\nu_\mu\to\nu_\mu$
experiments
and from
the 90\% CL exclusion plots of the
BNL E734
\cite{BNLE734},
BNL E776
\cite{BNLE776}
and
CCFR
\cite{CCFR96}
$
\stackrel{\makebox[0pt][l]
{$\hskip-3pt\scriptscriptstyle(-)$}}{\nu_{\mu}}
\to\stackrel{\makebox[0pt][l]
{$\hskip-3pt\scriptscriptstyle(-)$}}{\nu_{e}}
$
experiments.
The shadowed region
corresponds to the range (\ref{LSNDrange}) of $\Delta{m}^2$
allowed at 90\% CL by the results of the LSND
and all the other SBL experiments.
From Fig. \ref{fig4}
it can be seen that
$ |J_{e\mu}| \lesssim 10^{-1} $
for
$ \Delta{m}^2 \gtrsim 0.27 \, \mathrm{eV}^2 $,
which is an interval that includes the
LSND-allowed range (\ref{LSNDrange}).

For $|J_{\mu\tau}|$
we have the unitarity bound
\begin{equation}\label{Jmt1}
|J_{\mu\tau}|
\leq
f_2\!\left(a^{0}_{\mu}, y_2(a^0_\mu) \right) = 
a^{0}_{\mu} \sqrt{1-a^{0}_{\mu}}
\,.
\end{equation}
This limit is more stringent than
the corresponding one for
$|I_{\mu\tau}|$
given in Eq.(\ref{Imt1}).
Its numerical value
obtained from
the 90\% CL exclusion plots of the
CDHS \cite{CDHS84} and CCFR \cite{CCFR84}
$\nu_\mu\to\nu_\mu$
experiments
is shown by the solid curve in Fig. \ref{fig5}.
The dashed curve in Fig. \ref{fig5}
represents the value of the amplitude bound
\begin{equation}\label{Jmt2}
|J_{\mu\tau}|
\leq
\left\{
\begin{array}{lcl} \displaystyle
\frac{ 1 }{ 2 }
\sqrt{
A_{\mu;\tau}^{0}
\left(
4 \, a^{0}_{\mu}
-
A_{\mu;\tau}^{0}
\right)
}
& \quad \mbox{for} \quad &
A_{\mu;\tau}^{0}
\leq
2 \, a^{0}_{\mu}
\,,
\\[3mm] \displaystyle
a^{0}_{\mu}
& \quad \mbox{for} \quad &
A_{\mu;\tau}^{0}
\geq
2 \, a^{0}_{\mu}
\,,
\end{array} \right.
\end{equation}
obtained from
the 90\% CL exclusion plots of the
CDHS \cite{CDHS84} and CCFR \cite{CCFR84}
$\nu_\mu\to\nu_\mu$
experiments
and from
the 90\% exclusion plots
of the
FNAL E531
\cite{FNALE531}
and
CCFR
\cite{CCFR95}
$\nu_{\mu}\to\nu_{\tau}$
experiments.
From Fig. \ref{fig5}
one can see that
$ |J_{\mu\tau}| \lesssim 0.25 $
for
$ \Delta{m}^2 \gtrsim 0.3 \, \mathrm{eV}^2 $,
including the
LSND-allowed range (\ref{LSNDrange}),
and
$ |J_{\mu\tau}| \lesssim 8 \times 10^{-2} $
for
$ \Delta{m}^2 \gtrsim 0.5 \, \mathrm{eV}^2 $.

Since $1-c_e$ is large
and there is no constraint on the value of $c_\tau$
(and the available information on $A_{e;\tau}$
is rather poor),
the parameter
$|J_{e\tau}|$
is only subject to the unitarity bound
\begin{equation}
|J_{e\tau}|
\leq
f_2\!\left(1-a^{0}_{e}, y_2(1-a^0_e) \right) = 
\left( 1 - a^{0}_{e} \right) \sqrt{a^{0}_{e}}
\,.
\label{Jet1}
\end{equation}
This bound is much less stringent than the corresponding one for
$|I_{e\tau}|$,
represented by the upper
function in Eq.(\ref{Ime1})
and depicted as the solid curve in Fig. \ref{fig2}.
Since
$ a^{0}_e \lesssim 4 \times 10^{-2} $
for
$\Delta{m}^{2}$
in the wide range (\ref{widerange}),
we obtain the upper bound
$ |I_{e\tau}| \lesssim 0.2 $,
which is about half of the unitarity limit
$2/3\sqrt{3}\simeq0.385$.

From the bounds depicted in Figs. \ref{fig2} and \ref{fig4},
it is clear that the observation of a non-zero
T-odd asymmetry
$T_{\mu{e}}$
(and $\bar{T}_{\mu{e}}$)
in future LBL experiments
is a very difficult task.
On the other hand,
from Figs. \ref{fig3} and \ref{fig5}
one can see that the
T-odd asymmetry
$T_{\mu\tau}$
(and $\bar{T}_{\mu\tau}$)
could be rather large,
close to the maximal value
allowed by the unitarity of the neutrino mixing matrix.
Also the
T-odd asymmetry
$T_{e\tau}$
(and $\bar{T}_{e\tau}$)
could be rather large,
but not more than half of the maximal value
allowed by the unitarity of the neutrino mixing matrix
and
only if the matter effect is important and enhances
the contribution of
$J'_{e\tau}$.
Hence,
we conclude that
long-baseline
$ \nu_\mu \leftrightarrows \nu_\tau $
(and
$ \bar\nu_\mu \leftrightarrows \bar\nu_\tau $)
experiments
are favoured for the investigation
of CP (and T) violation in the lepton sector
if the four-neutrino schemes (\ref{AB})
are realized in nature.

We would also like to mention that
the CP-odd parameters
$J_{\alpha\beta}$
are relevant for the CP-odd asymmetries that could be measured
by extremely long-baseline (ELBL) oscillation experiments
with neutrino beams propagating in vacuum,
for which
$ \Delta{m}^{2}_{21} L / 2 p \gg 2\pi $
and
$ \Delta{m}^{2}_{43} L / 2 p \sim 1 $
(in scheme A),
\begin{equation}
D^{(\mathrm{ELBL})}_{\alpha;\beta}
=
J_{\alpha\beta}
\,
\sin
\frac{ \Delta{m}^{2}_{43}L }{ 2p }
\,,
\label{DELBL}
\end{equation}
although we do not know if it will ever be possible
to make such experiments.

Concluding this Section,
we briefly discuss the matter effects in the case
of the 3-neutrino schemes
considered in Section \ref{Three massive neutrinos}.
We consider,
for simplicity,
only scheme I,
but analogous conclusions are valid in scheme II.
The 3-neutrino scheme I
corresponds to the 4-neutrino scheme A,
with the difference that the solar sector
of the mixing matrix is absent,
i.e.
$U_{\alpha4}=0$
for
$\alpha=e,\mu,\tau$
and
$R_{\mathrm{sun}}=1$
(apart from negligible corrections
of order $a_{CC}/\Delta{m}^2$).
Hence,
in the 3-neutrino scheme I
the neutrino T-odd asymmetries $T_{\alpha\beta}$
(see Eq.(\ref{Todd}))
are given by Eq.(\ref{Tnu}) 
with
$J'_{\alpha\beta}=0$.
Using the same reasoning as that employed in the case
of the 4-neutrino schemes,
one can see that the upper bounds on
$I_{\alpha\beta}$
derived in Section \ref{Three massive neutrinos}
in the case of the 3-neutrino schemes
are also valid for
$I'_{\alpha\beta}$
and
$\bar{I}'_{\alpha\beta}$.
Hence,
$T_{e\mu}$
is very small
if the neutrino mixing parameters lie in region 1
and
is less suppressed in regions 2 and 3.

\section{Conclusions}
\label{Conclusions}

In this paper
we have considered
possibilities to reveal effects of CP-violation in the lepton sector
in future accelerator
long-baseline (LBL) experiments
(K2K \cite{K2K}, MINOS \cite{MINOS}, ICARUS \cite{ICARUS}
and others \cite{otherLBL}).

At present there are three experimental indications
in favour of neutrino
oscillations which correspond to three different scales of
neutrino mass-squared differences:
the solar neutrino deficit, the atmospheric neutrino
anomaly and the result of the LSND experiment.
These indications
and the negative results of numerous short-baseline (SBL) neutrino
experiments can be accommodated in the two four-neutrino
schemes A and B presented in Eq.(\ref{AB})
\cite{BGG96A,BGG96B,OY96}.

Working in the framework of the neutrino mixing schemes A and B,
we have
derived constraints on the
parameters $I_{\alpha\beta}$ ($\alpha, \beta = e, \mu, \tau$)
(see Eq.(\ref{iab}))
that characterize
CP violation in
$
\stackrel{\scriptscriptstyle(-)}{\nu}_{{\hskip-4pt}\alpha} 
\to
\stackrel{\scriptscriptstyle(-)}{\nu}_{{\hskip-4pt}\beta}
$
LBL neutrino oscillation experiments in vacuum.
These parameters are given in terms of the quantities
$I_{\alpha\beta;jk}$ which appear in
the general CP-odd asymmetries
(see Eqs.(\ref{061}) and (\ref{I}))
as $I_{\alpha\beta}=I_{\alpha\beta;12}$ in
scheme A and
$I_{\alpha\beta}=I_{\alpha\beta;34}$ in scheme B.
We have developed methods for 
deriving upper bounds on the
parameters $I_{\alpha \beta}$
from the data of
SBL experiments
which can be applied not only to the schemes A and B but to
arbitrary schemes with any number of neutrinos.
We have shown
that the CP-odd parameter
$I_{e\mu}$
is bounded by
$| I_{e\mu} | \lesssim 10^{-2}$
(see Fig. \ref{fig2})
and
a similar suppression applies to
$I_{e\tau}$.
On the other hand,
sizable CP violation can be expected in
$\nu_\mu \rightarrow \nu_\tau$
oscillations.
The CP-odd parameter relative to this channel
could be close to its maximally allowed
value
$|I_{\mu\tau}|_{\mathrm{max}} = 2/3\sqrt{3} \approx 0.385$, 
resulting from the unitarity of the mixing matrix (see Fig. \ref{fig3}).

For LBL accelerator experiments the matter background
is important. In this case the parameters
$I_{\alpha\beta;jk}$ for
neutrinos (antineutrinos) are replaced by
the CP-violating parameters $I'_{\alpha\beta;jk}$
($\bar{I}'_{\alpha\beta;jk}$)
which include matter effects
(see Eq.(\ref{I'})).
Using a physically motivated approximate
method of incorporating matter effects \cite{BGG97},
we have
demonstrated that the
quantities $|I'_{\alpha\beta}|$ ($|\bar{I}'_{\alpha\beta}|$) are
bounded by the same functions of the SBL mass-squared difference
$\Delta{m}^2$ as $|I_{\alpha\beta}|$ apart from terms of order
$a_{CC}/\Delta{m}^2$.
Therefore, the bounds
depicted in Figs. \ref{fig2} and \ref{fig3} apply
also to the corresponding
$I'_{\alpha\beta}$ and $\bar{I}'_{\alpha\beta}$.
However, although finding
$I'_{\alpha\beta} \neq 0$ and/or $\bar{I}'_{\alpha\beta} \neq 0$
would prove the existence of CP violation
in the neutrino mixing matrix $U$, the knowledge of the
parameters
$I'_{\alpha\beta}$ and/or $\bar{I}'_{\alpha\beta}$
cannot easily be transformed into
information on $I_{\alpha\beta}$ and thus $U$, because both
sets of parameters are related in complicated, non-linear way
involving also the ``matter potentials'' $a_{CC}$, $a_{NC}$ and
the mass-squared difference relevant for LBL neutrino oscillations.

We have also shown that in matter a second CP-violating parameter
$J'_{\alpha\beta}$
(given by $J'_{\alpha\beta}=I'_{\alpha\beta;34}$ in scheme A
and $J'_{\alpha\beta}=I'_{\alpha\beta;12}$ in scheme B)
appears in the $\nu_\alpha\to\nu_\beta$ transition
probability.
Its contribution is
only significant if the oscillation phase $\omega$
defined in Eq.(\ref{po})
is sufficiently large.
An evaluation of the maximal value
that the phase $\omega$ can assume
in accelerator LBL experiments
shows that the contribution of
$J'_{\alpha\beta}$
could be relevant for baselines longer than
$ \sim 100 \, \mathrm{km} $
(see Eq.(\ref{omax})).
We have argued that the parameter
$J'_{\alpha\beta}$
(and the analogous parameter
$\bar{J}'_{\alpha\beta}$
for antineutrinos)
is subject to the same bounds as
$J_{\alpha\beta}$
(with
$J_{\alpha\beta} \equiv I_{\alpha\beta;34}$
in scheme A
and
$J_{\alpha\beta} \equiv I_{\alpha\beta;12}$
in scheme B).
These bounds are presented in Eqs.(\ref{Jme1})--(\ref{Jmt2})
and their numerical values
obtained from the results of disappearance and appearance
SBL neutrino oscillation experiments are shown by
the curves in Figs. \ref{fig4} and \ref{fig5}. 
There is no analogue of the parameters $J'_{\alpha\beta}$
and $\bar{J}'_{\alpha\beta}$ in the 3-neutrino case.

Since a measurement
of the CP-odd asymmetries
$D_{\alpha;\beta}$
defined in Eq.(\ref{Dab})
does not allow to obtain direct information
on CP violation
if matter effects are important,
we have considered
the long-baseline T-odd asymmetries
$T_{\alpha;\beta}$ ($\bar{T}_{\alpha;\beta}$)
defined in Eq.(\ref{Todd}).
Since the matter contribution
to the effective neutrino and antineutrino
Hamiltonians (\ref{Hnu}), (\ref{Hanu})
is real
and
the matter density is symmetric along the path
of the neutrino beam
in terrestrial LBL experiments
(to a good approximation it is even constant for baselines shorter than
$ \sim 4000 \, \mathrm{km} $)
the matter effects are T-symmetric.
Therefore,
the T-odd asymmetries
$T_{\alpha;\beta}$ ($\bar{T}_{\alpha;\beta}$)
are only different from zero if CP is violated in the lepton mixing matrix
$U$.
We have shown that
in the four-neutrino schemes A and B
the T-odd asymmetries
depend
on the parameters
$I'_{\alpha\beta}$ ($\bar{I}'_{\alpha\beta}$) and
$J'_{\alpha\beta}$ ($\bar{J}'_{\alpha\beta}$)
(see Eq.(\ref{Tnu})).
Hence,
they are subject to the constraints
derived from the results of SBL experiments.

In conclusion,
we have shown that
in the four-neutrino schemes A and B
the channels
$
\stackrel{\scriptscriptstyle(-)}{\nu}_{{\hskip-4pt}\mu} 
\leftrightarrows
\stackrel{\scriptscriptstyle(-)}{\nu}_{{\hskip-4pt}e}
$
are disfavoured for the search of
CP-violation effects in future
LBL oscillation experiments,
the channels
$
\stackrel{\scriptscriptstyle(-)}{\nu}_{{\hskip-4pt}e} 
\leftrightarrows
\stackrel{\scriptscriptstyle(-)}{\nu}_{{\hskip-4pt}\tau}
$
could allow to reveal relatively large
CP-violating effects
where matter plays an important role
and
the channels
$
\stackrel{\scriptscriptstyle(-)}{\nu}_{{\hskip-4pt}\mu} 
\leftrightarrows
\stackrel{\scriptscriptstyle(-)}{\nu}_{{\hskip-4pt}\tau}
$
could show
CP-violating effects 
as large as is allowed by the unitarity
of the neutrino mixing matrix.

%\acknowledgments

\newpage
\appendix

\section{Derivation of the amplitude bound}
\label{apa}

In this appendix we discuss
the derivation of the ``amplitude bound''.
The starting point is the quantity
\begin{equation}
I_{\alpha\beta}
=
4
\,
\mbox{Im}\!\left[
U_{\alpha1} \, U_{\beta1}^* \, U_{\alpha2}^* \, U_{\beta2}
\right]
\qquad (\alpha \neq \beta)
\,,
\end{equation}
which determines the CP-odd asymmetry
in the four-neutrino scheme A.
The same bound can be derived in the scheme B
with the change of indices (\ref{0721}).

It is obvious that $I_{\alpha\beta}$ is invariant under the phase 
transformations
\begin{equation}
U_{\alpha j} \rightarrow \mbox{e}^{i \gamma_j} U_{\alpha j}\,,
\qquad
U_{\beta j}  \rightarrow \mbox{e}^{i \gamma_j}  U_{\beta j}
\,, 
\end{equation}
where the $\gamma_j$ are arbitrary phases.
Thus the elements $U_{\alpha j}$
can be taken to be real.
Taking into account
the definition (\ref{dc}),
we can write
\begin{equation}
U_{\alpha j} = \sqrt{c_\alpha} \, e^{(1)}_j
\quad
\mbox{with}
\quad
j=1,2
\label{Ual}
\end{equation}
and the orthonormal basis
\begin{equation}\label{ON}
e^{(1)}(\theta) = (\cos \theta, \sin \theta)
\,,
\qquad
e^{(2)}(\theta) = (-\sin \theta, \cos \theta)
\,.
\end{equation}
We expand $U_{\beta j}$
with respect to this
basis as
\begin{equation}\label{Ube}
U_{\beta j}
=
\sqrt{c_\beta} \sum_{\rho=1,2} p_\rho \, e^{(\rho)}_j
\,, 
\end{equation}
where
$p_1$ and $p_2$
are complex coefficients
such that 
\begin{equation}\label{norm}
\sum_{\rho=1,2} |p_\rho|^2 = 1
\,.
\end{equation}
With the help of
Eqs.(\ref{Ual})--(\ref{norm})
we easily find
\begin{equation}\label{Kp}
I_{\alpha\beta}
=
2 \, c_\alpha \, c_\beta \, \sin 2\theta \, 
\mbox{Im}(p_1^* p_2) =
2 \, c_\alpha \, c_\beta
\, |p_1| \, \sqrt{1-|p_1|^2}
\, \sin 2\theta \, \sin \delta
\,,
\end{equation}
where $\delta$ is the phase of $p_1^* p_2$.

The parameter $|p_1|$ is connected to the oscillation amplitude
$A_{\alpha;\beta}$ and the parameters $c_\alpha$, $c_\beta$.
In fact,
from Eqs.(\ref{Ual}) and (\ref{Ube}) we have
\begin{equation}
A_{\alpha;\beta}
=
4 \left| \sum_{j=1,2} U_{\alpha j} U_{\beta j}^* \right|^2
=
4 \, c_\alpha \, c_\beta \, |p_1|^2 \,,
\end{equation}
which implies
$ |p_1| = \sqrt{ A_{\alpha;\beta} / 4 c_\alpha c_\beta } $.
Inserting this expression in
Eq.(\ref{Kp}),
we obtain
\begin{equation}
I_{\alpha\beta} =
\frac{1}{2}
\,
\sqrt{
A_{\alpha;\beta}
\left(
4 \, c_{\alpha} \, c_{\beta}
-
A_{\alpha;\beta}
\right)
}
\, \sin 2\theta \, \sin \delta
\end{equation}
and thus we arrive at the ``amplitude bound''
\begin{equation}
|I_{\alpha\beta}|
\leq
\frac{1}{2} \,
\sqrt{
A_{\alpha;\beta} 
\left( 
4 \, c_{\alpha} \, c_{\beta} - A_{\alpha;\beta}
\right)
}
\,.
\label{oab}
\end{equation}
Let us stress that this derivation is based only on the obvious
inequality
\begin{equation}
| \sin 2\theta \sin \delta | \leq 1
\,.
\end{equation}
Since $c_\alpha$, $c_\beta$ and
$A_{\alpha;\beta}$ do not restrict $\theta$ and $\delta$,
the bound
(\ref{oab}) is the optimal one.

\section{Derivation of the unitarity bound}
\label{apb}

Up to now we did not use
the unitarity of the mixing matrix.
Taking this fact
into account will allow us to obtain an upper bound on
$|I_{\alpha\beta}|$
depending solely on $c_\alpha$ and $c_\beta$.

The unitarity of the mixing matrix tells us that
\begin{equation}
\sum_{j=1,2} U_{\alpha j} U_{\beta j}^*
=
-
\sum_{j=3,4} U_{\alpha j} U_{\beta j}^*
\,.
\end{equation}
This relation allows to write the
oscillation amplitude
$A_{\alpha;\beta}$
in the two forms of Eq.(\ref{A12}).
Using the Cauchy--Schwarz inequality,
one can see that
(in the scheme A)
\begin{equation}
A_{\alpha;\beta}
=
4
\left| \sum_{j=3,4} U_{\alpha j} U_{\beta j}^* \right|^2
\leq
4
\left( \sum_{j=3,4} |U_{\alpha j}|^2 \right)
\left( \sum_{j=3,4} |U_{\beta j}|^2 \right)
=
4 \, (1-c_\alpha) \, (1-c_\beta)
\,.
\label{cs}
\end{equation}

The right-hand side of the inequality
(\ref{oab}),
as a function of $A_{\alpha;\beta}$,
reaches its maximum,
$ c_\alpha c_\beta $,
at
(here we do not take into account possible experimental information
on $A_{\alpha;\beta}$)
\begin{equation}
(A_{\alpha;\beta})_0
=
2 \, c_\alpha \, c_\beta
\,. 
\end{equation}
Consequently, if the condition
\begin{equation}\label{ineq}
2 \, (1-c_\alpha) \, (1-c_\beta)
\geq
c_\alpha \, c_\beta
\end{equation}
is satisfied,
the upper bound (\ref{cs}) on
$A_{\alpha;\beta}$
is larger then
$(A_{\alpha;\beta})_0$.
In this case we have
\begin{equation}
|I_{\alpha\beta}| \leq c_\alpha \, c_\beta
\,.
\end{equation}
If the condition (\ref{ineq}) is not fulfilled,
the upper bound
(\ref{cs}) is smaller than $(A_{\alpha;\beta})_0$
and has to be
inserted for $A_{\alpha;\beta}$
into Eq.(\ref{oab}),
leading to
\begin{equation}
|I_{\alpha\beta}| \leq
2 \,
\sqrt{
(c_\alpha + c_\beta -1) \, (1-c_\alpha) \, (1-c_\beta)
}
\,.
\label{B6}
\end{equation}
Thus, we arrive at the ``unitarity bound''
\begin{equation}
|I_{\alpha\beta}| \leq f(c_\alpha , c_\beta)
\,,
\end{equation}
with the function 
\begin{equation}\label{f}
f(x,y) = 
\left\{ \begin{array}{lcl} \displaystyle 
f_1 \equiv xy
& \quad \mbox{for} \quad &
2(1-x)(1-y) \ge xy
\,,
\\[3mm] \displaystyle
f_2 \equiv 2 [(x+y-1)(1-x)(1-y)]^{1/2}
& \quad \mbox{for} \quad &
2(1-x)(1-y) < xy
\,,
\end{array} \right.
\end{equation}
defined on the unit square $0 \le x \le 1$, $0 \le y \le 1$.
The
function
\begin{equation}
g(x) = \frac{2\,(1-x)}{2-x}
\end{equation}
represents the borderline separating the two regions in the
definition of the function (\ref{f}).
It is clear from our
derivation (and also easy to check)
that $f$
is continuous along this borderline.

\section{Discussion of the function $\lowercase{f}$}
\label{apc}

Since we do not have
definite experimental values of
$c_\alpha$
and
$c_\beta$,
but only bounds on these quantities
(see Eqs.(\ref{A}) and (\ref{B})),
which
define allowed rectangles in the square
$0 \leq c_\alpha \leq 1$,
$0 \leq c_\beta \leq 1$,
we are interested in the behaviour of $f$
in order to evaluate the
unitarity bound.

From the partial derivative of $f$
in the region $y \ge g(x)$, 
\begin{equation}\label{part}
\frac{\partial f}{\partial x} = \frac{\partial f_2}{\partial x}
\propto (2-2x-y)
\,,
\end{equation}
one can see that,
at fixed $y$,
the function $f$ increases monotonously
from $x=0$ to the point $x=1-y/2$,
where the partial derivative in Eq.(\ref{part}) is zero.
The points
$x=1-y/2$ lie on the straight line $y_1(x)=2-2x$.
In the range
$1-y/2 \leq x \leq 1$
the function $f$ decreases monotonously.
Taking
into account the symmetry $f(x,y)=f(y,x)$,
we see that at fixed $x$ the function $f$
increases monotonously from $y=0$ to the point $y=1-x/2$,
where the
partial derivative of $f$ with respect to $y$ is zero.
These points
lie on the straight line $y_2(x)=1-x/2$.
Beyond this line $f$
decreases monotonously. Note that both straight lines lie in the range
of $f_2$.

Figure \ref{fig1} shows a contour plot of
the function $f(x,y)$,
together with the lines
$y_1$ and $y_2$ which intersect at the point
\begin{equation}\label{point}
x = y = \frac{2}{3}
\,.
\end{equation}
At this point both partial derivatives of $f$
are equal to zero and 
therefore the point (\ref{point})
corresponds to the absolute maximum of $f$,
given by
\begin{equation}
f_{\mbox{\scriptsize max}}
=
f_2\!\left(\frac{2}{3},\frac{2}{3}\right)
=
\frac{2}{3^{3/2}} 
\approx
0.385
\,.
\end{equation}
This number constitutes the absolute
upper bound for $|I_{\alpha\beta}|$.

\begin{figure}[h]
\refstepcounter{figure}
\label{fig1}
FIG.\ref{fig1}.
Contour plot of the function
$f(x,y)$ given in Eq.(\ref{fxy}).
The dotted line
is the borderline
$g(x)=2(1-x)/(2-x)$
between the regions where
$f=f_{1}$
and
$f=f_{2}$.
The two solid lines represent the functions
$y_1(x)=2-2x$
and 
$y_2(x)=1-x/2$.
\end{figure}

\begin{figure}[h]
\refstepcounter{figure}
\label{fig2}
FIG.\ref{fig2}.
Upper bound for
the parameter
$|I_{e\mu}|$
which characterizes the CP-odd asymmetry
in the $\nu_{\mu}\to\nu_{e}$ channel
for the SBL parameter
$\Delta{m}^2$
in the range
$
10^{-1} \, \mathrm{eV}^2
\leq \Delta{m}^2 \leq
10^{3} \, \mathrm{eV}^2
$.
The solid curve
represents the upper function
in Eq.(\ref{Ime1})
and is obtained 
from the 90\% CL exclusion plot of the Bugey
$\bar\nu_e\to\bar\nu_e$ experiment.
The dash-dotted curve improves the solid curve
where $a^0_\mu \leq a^0_e/2$
(the lower function in Eq.(\ref{Ime1})).
It is obtained
from the 90\% CL exclusion plots of the Bugey
$\bar\nu_e\to\bar\nu_e$ experiment
and the CDHS and CCFR
$\nu_{\mu}\to\nu_{\mu}$
experiments.
The dashed curve is obtained 
from the 90\% CL exclusion plots of the Bugey
$\bar\nu_e\to\bar\nu_e$ experiment
and the
BNL E734,
BNL E776 and
CCFR
$\nu_\mu\to\nu_e$
and
$\bar\nu_\mu\to\bar\nu_e$
experiments
(see the upper function in  Eq.(\ref{Ime2})).
The shadowed region
corresponds to the range (\ref{LSNDrange}) of $\Delta{m}^2$
allowed at 90\% CL by the results of the LSND experiment.
The solid curve represents also an upper bound for
$|I_{e\tau}|$.
\end{figure}

\begin{figure}[h]
\refstepcounter{figure}
\label{fig3}
FIG.\ref{fig3}.
Upper bound for
the parameter $|I_{\mu\tau}|$
which characterizes the CP-odd asymmetry
in the $\nu_{\mu}\to\nu_{\tau}$ channel.
The solid curve is obtained 
from the 90\% CL exclusion plots of the CDHS and CCFR
$\nu_{\mu}\to\nu_{\mu}$
experiments
(see Eq.(\ref{Imt1})).
The dashed curve is obtained 
from the 90\% CL exclusion plots of the FNAL E531 and CCFR
$\nu_\mu\to\nu_\tau$
experiments
(see Eq.(\ref{Imt2})).
The shadowed region
corresponds to the range (\ref{LSNDrange}) of $\Delta{m}^2$
allowed at 90\% CL by the results of the LSND experiment.
\end{figure}

\begin{figure}[h]
\refstepcounter{figure}
\label{fig4}
FIG.\ref{fig4}.
Upper bound for
the parameter
$|J_{e\mu}|$
(see Eq.(\ref{Jab})).
The solid curve
represents the unitarity bound (\ref{Jme1})
and is obtained 
from the 90\% CL exclusion plots of the Bugey
$\bar\nu_e\to\bar\nu_e$ experiment
and those of
the CDHS and CCFR
$\nu_{\mu}\to\nu_{\mu}$
experiments.
The dashed curve
represents the value of the amplitude bound (\ref{Jme2})
obtained 
from the 90\% CL exclusion plots of
the CDHS and CCFR
$\nu_{\mu}\to\nu_{\mu}$
experiments
and those of the
BNL E734,
BNL E776 and
CCFR
$\nu_\mu\to\nu_e$
and
$\bar\nu_\mu\to\bar\nu_e$
experiments.
The shadowed region
corresponds to the range (\ref{LSNDrange}) of $\Delta{m}^2$
allowed at 90\% CL by the results of the LSND experiment.
\end{figure}

\begin{figure}[h]
\refstepcounter{figure}
\label{fig5}
FIG.\ref{fig5}.
Upper bound for
the parameter $|J_{\mu\tau}|$
(see Eq.(\ref{Jab})).
The solid curve represents the unitarity bound (\ref{Jmt1})
obtained 
from the 90\% CL exclusion plots of the CDHS and CCFR
$\nu_{\mu}\to\nu_{\mu}$
experiments.
The dashed curve
depicts the value of the amplitude bound (\ref{Jmt2})
obtained 
from the 90\% CL exclusion plots of the CDHS and CCFR
$\nu_{\mu}\to\nu_{\mu}$
experiments
and those of the FNAL E531 and CCFR
$\nu_\mu\to\nu_\tau$
experiments.
The shadowed region
corresponds to the range (\ref{LSNDrange}) of $\Delta{m}^2$
allowed at 90\% CL by the results of the LSND experiment.
\end{figure}

%\end{document}

\newpage

\begin{minipage}[p]{0.95\textwidth}
\begin{center}
\mbox{\epsfig{file=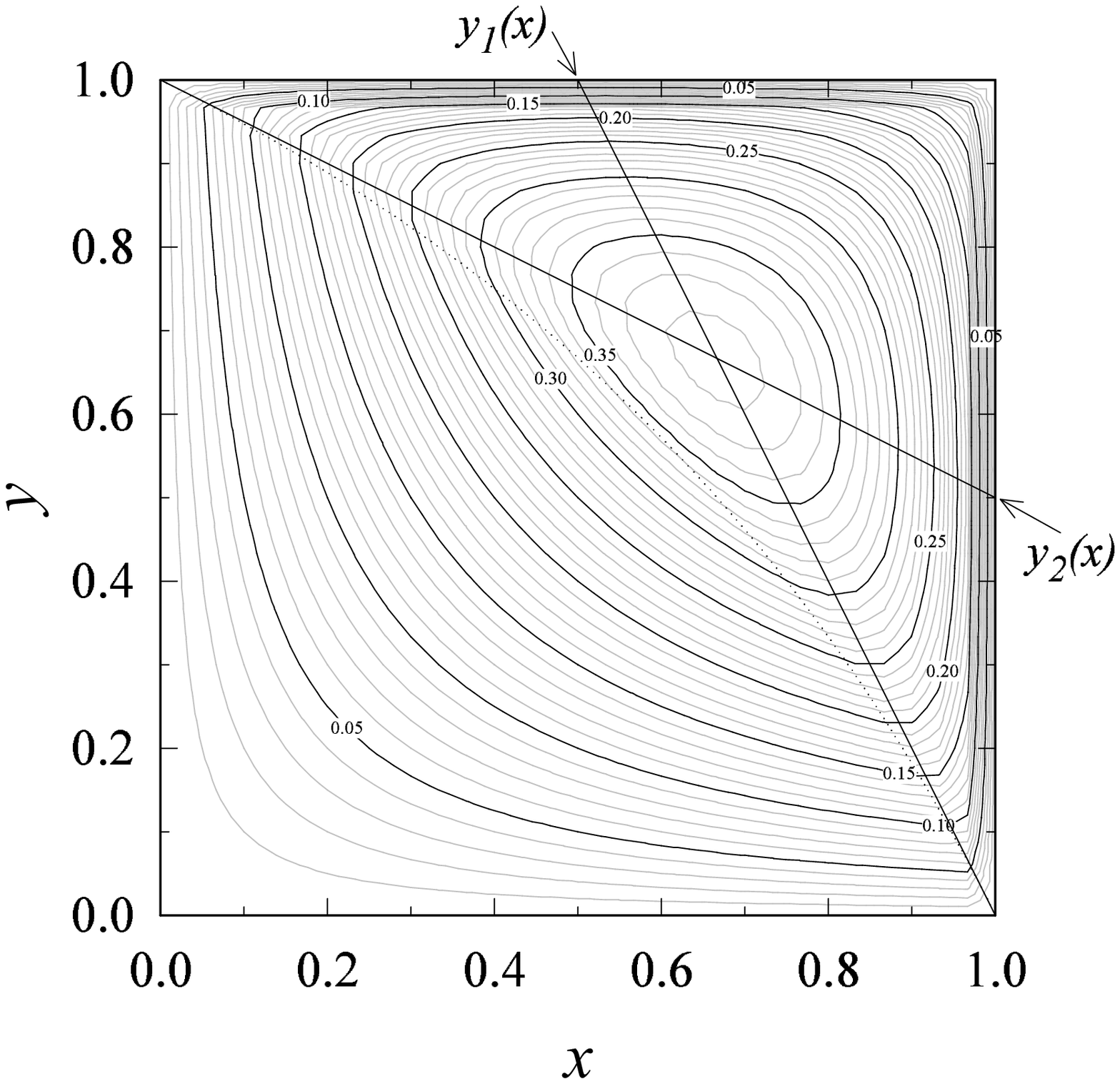,height=0.95\textheight}}
\end{center}
\end{minipage}
\begin{center}
\Large Figure~\ref{fig1}
\end{center}

\newpage

\begin{minipage}[p]{0.95\textwidth}
\begin{center}
\mbox{\epsfig{file=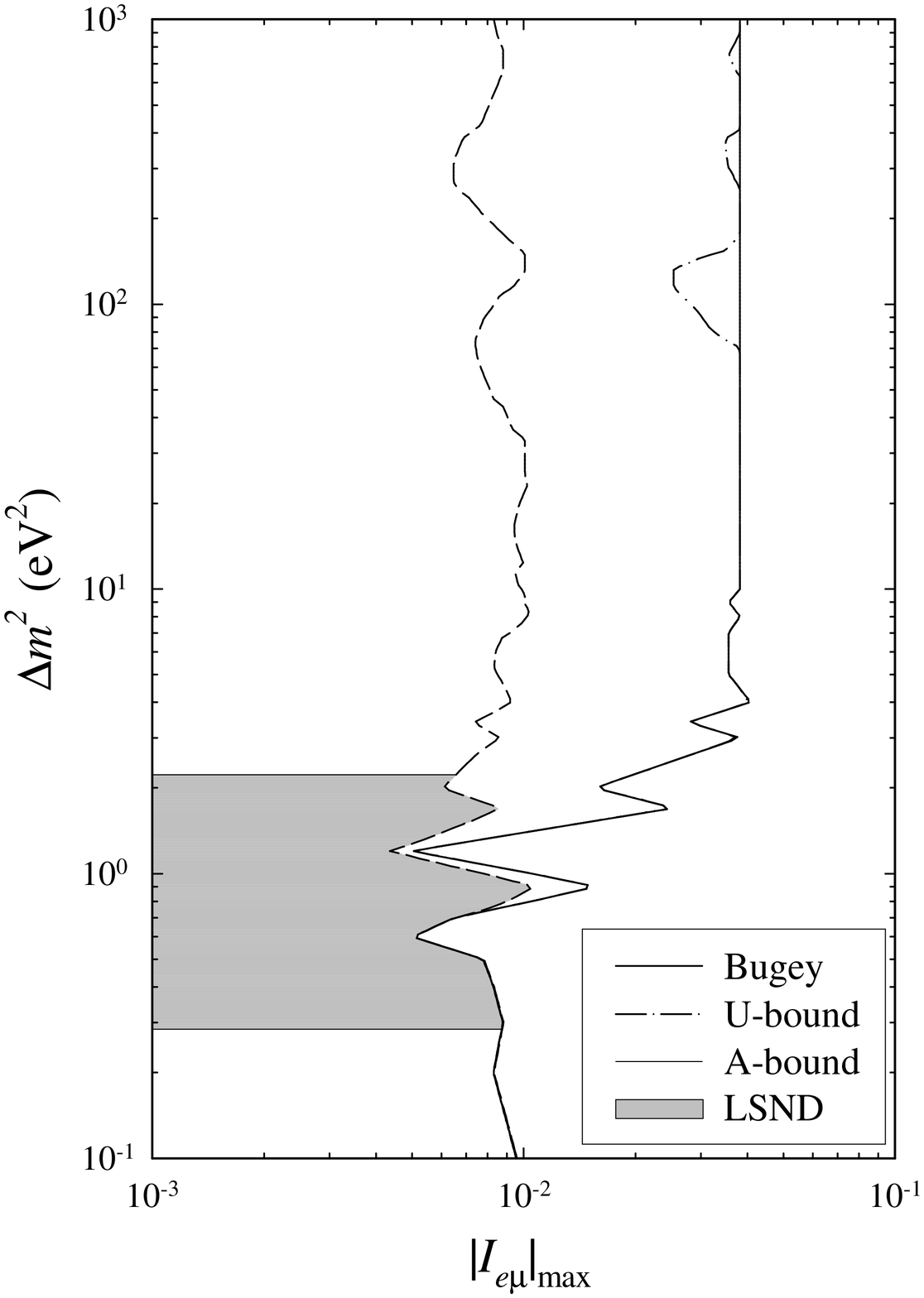,height=0.95\textheight}}
\end{center}
\end{minipage}
\begin{center}
\Large Figure~\ref{fig2}
\end{center}

\newpage

\begin{minipage}[p]{0.95\textwidth}
\begin{center}
\mbox{\epsfig{file=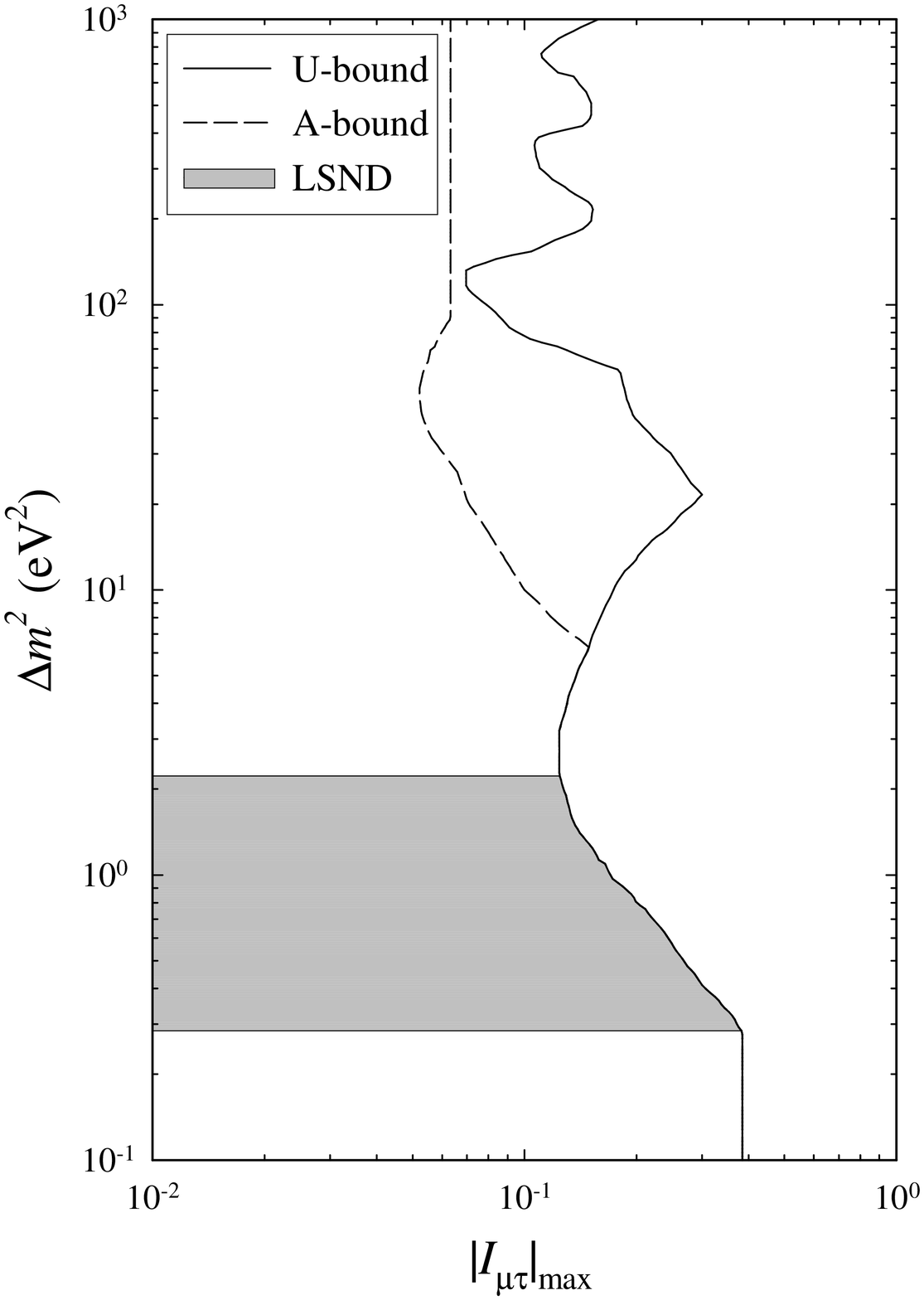,height=0.95\textheight}}
\end{center}
\end{minipage}
\begin{center}
\Large Figure~\ref{fig3}
\end{center}

\newpage

\begin{minipage}[p]{0.95\textwidth}
\begin{center}
\mbox{\epsfig{file=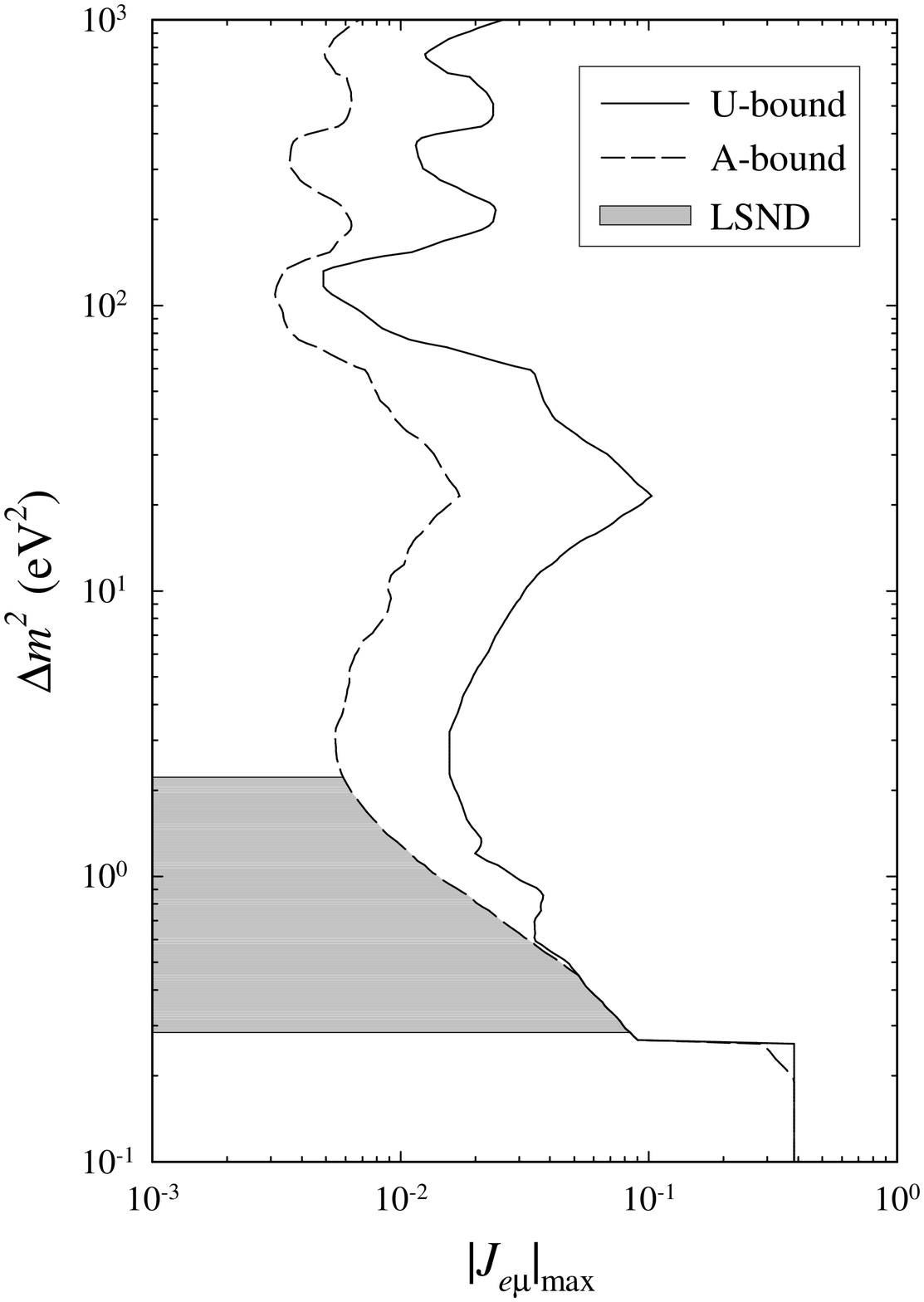,height=0.95\textheight}}
\end{center}
\end{minipage}
\begin{center}
\Large Figure~\ref{fig4}
\end{center}

\newpage

\begin{minipage}[p]{0.95\textwidth}
\begin{center}
\mbox{\epsfig{file=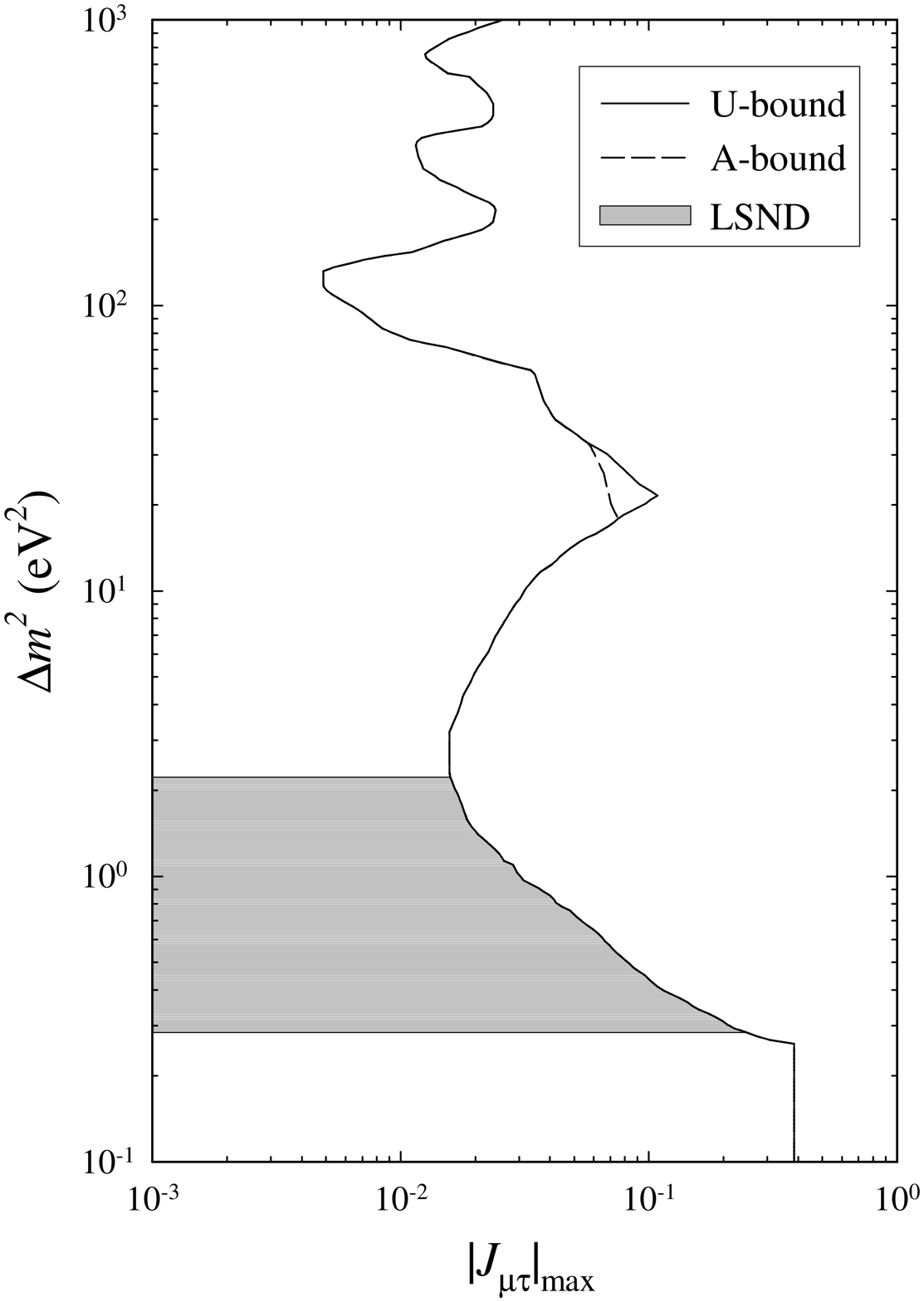,height=0.95\textheight}}
\end{center}
\end{minipage}
\begin{center}
\Large Figure~\ref{fig5}
\end{center}

\end{document}